\documentclass[12pt]{article}

\usepackage[utf8]{inputenc}
\usepackage{longtable, ltcaption}%
\usepackage{amsmath,amsthm,bbm}
\usepackage[margin=1in]{geometry}
\usepackage{setspace,placeins,booktabs}
\usepackage{hyperref}
\usepackage{cleveref, graphicx, caption, xcolor, multirow}
\usepackage{dcolumn, ctable, subcaption}
\usepackage[bibstyle=numeric, citestyle=authoryear, doi=false, url=true, backend=biber, maxbibnames=10, maxcitenames=4, uniquelist=false, uniquename=false, sorting=nyt]{biblatex}
\renewbibmacro{in:}{}
\DeclareNameAlias{default}{last-first/first-last}
\newcommand{\citet}{\textcite}
\newcommand{\indicator}[1]{\mathbbm{1}\{#1\}}
\newcommand\independent{\protect\mathpalette{\protect\independenT}{\perp}}
    \def\independenT#1#2{\mathrel{\setbox0\hbox{$#1#2$}%
    \copy0\kern-\wd0\mkern4mu\box0}}

\newtheorem{assumption}{Assumption}

\crefname{figure}{Figure}{Figures}
\crefname{assumption}{Assumption}{Assumptions}
\crefname{inneruassumption}{Assumption}{Assumptions}

\theoremstyle{definition}
\newtheorem{algorithm}{Algorithm}

\newtheorem{remark}{Remark}

\begin{document}
\title{Heterogeneous Effects of Job Displacement on Earnings}

\date{September 30, 2020}
\author{Afrouz Azadikhah Jahromi\footnote{Assistant Professor, Department of Economics, Widener University, Email: aazadikhahjahromi@widener.edu.}  \and Brantly Callaway\footnote{Assistant Professor, Department of Economics, University of Georgia, Email: brantly.callaway@uga.edu}\footnote{We thank Guest Editor Blaise Melly and an anonymous referee for a number of helpful comments.  We also would like to thank Michael Bognanno, Bernd Fitzenberger, Weige Huang, Moritz Ritter, Jungmo Yoon, and participants at the Economic Applications of Quantile Regression 2.0 Conference for their comments and suggestions.} }
              
\maketitle

\begin{abstract}
  \noindent This paper considers how the effect of job displacement varies across different individuals.  In particular, our interest centers on features of the distribution of the \textit{individual-level} effect of job displacement.  Identifying features of this distribution is particularly challenging -- e.g., even if we could randomly assign workers to be displaced or not, many of the parameters that we consider would not be point identified.  We exploit our access to panel data, and our approach relies on comparing outcomes of displaced workers to outcomes the same workers would have experienced if they had not been displaced and if they maintained the same rank in the distribution of earnings as they had before they were displaced. Using data from the Displaced Workers Survey, we find that displaced workers earn about \$157 per week less, on average, than they would have earned if they had not been displaced.  We also find that there is substantial heterogeneity.  We estimate that 42\% of workers have higher earnings than they would have had if they had not been displaced and that a large fraction of workers have experienced substantially more negative effects than the average effect of displacement.   Finally, we also document major differences in the distribution of the effect of job displacement across education levels, sex, age, and counterfactual earnings levels.  Throughout the paper, we rely heavily on quantile regression.  First, we use quantile regression as a flexible (yet feasible) first step estimator of conditional distributions and quantile functions that our main results build on.  We also use quantile regression to study how covariates affect the distribution of the individual-level effect of job displacement.  
\end{abstract}

\bigskip


\bigskip


\bigskip

\vspace{100pt}

\pagebreak

\normalsize
\onehalfspacing

\section{Introduction}

A typical finding in the job displacement literature is that ``the effect'' of job displacement is large and persistent (see, for example, \citet{jacobson-lalonde-sullivan-1993} and much subsequent work).  However, considering a single effect of job displacement ignores potential heterogeneity across individuals.  Treatment effect heterogeneity is a general feature of many program evaluation problems in economics, but treatment effect heterogeneity is likely to be particularly important in the context of job displacement.  In particular, some individuals may be able to find similar jobs quickly after they are displaced, but other individuals may experience larger losses due to decreased match quality with their employer (e.g., \citet{jovanovic-1979}), loss of firm-, occupation-, or industry-specific human capital (\citet{topel-1991,neal-1995}), loss of a job at a high paying firm (\citet{abowd-kramarz-margolis-1999}), or failure to find full time work (\citet{farber-2017}).\footnote{See \citet{carrington-fallick-2017} for a good discussion of reasons for why earnings may decline following job displacement.}  In the current paper, we are interested in trying to distinguish between the cases where all individuals experience roughly the same effect of job displacement or where ``the effect''of job displacement mixes together individuals who have relatively small effects of job displacement with individuals who experience large negative effects of job displacement.  Distinguishing between these cases is important for our understanding of job displacement as well as being useful for targeting policy responses to job displacement.

Learning about treatment effect heterogeneity requires learning about the distribution of the treatment effect itself.\footnote{A related (though distinct) idea is to compare the distribution of observed outcomes for displaced workers to their counterfactual distribution of outcomes if they had not been displaced.  Researchers often compare the quantiles of these two distributions -- this class of parameters is called quantile treatment effects.  However, these types of parameters are likely to have severe limitations for understanding heterogeneous effects of job displacement which we discuss in more detail in the next section.}  And the distribution of the treatment effect itself depends, in turn, on the joint distribution of treated and untreated potential outcomes.  Learning about the joint distribution of treated and untreated potential outcomes is quite challenging -- e.g., even if we could randomly assign workers' displacement status, this would only identify the marginal distributions not the joint -- and has been the subject of a large literature in econometrics.  This literature includes \citet{heckman-smith-clements-1997,fan-park-2010,fan-guerre-zhu-2017,frandsen-lefgren-2017,callaway-2020}, among others.  This literature has largely focused on partial identification results under relatively weak assumptions or invoking stronger assumptions to obtain point identification.

In the current paper, we identify the joint distribution of earnings for displaced workers (their treated potential outcomes) and the outcomes they would have experienced if (i) they had not been displaced from their job and (ii) they maintained the same rank in the earnings distribution as they had before they were displaced.  Under the additional assumption that displaced workers would have maintained their rank in the earnings distribution over time (we call this assumption rank invariance over time), the joint distribution of treated and untreated potential outcomes is point identified and the above effects are due to job displacement.  %

The first step to understanding heterogeneous effects of job displacement is to estimate the (conditional) distributions of treated and untreated potential outcomes.  This is a non-trivial task (though one we have largely abstracted from so far because these sorts of arguments are more standard).  For identification, we use the approach of \citet{callaway-li-oka-2018}.  The two main identifying assumptions are that the distribution of the change in untreated potential outcomes over time is independent of treatment status after conditioning on covariates (this is a Difference in Differences-type assumption) and that the conditional copula for the change in untreated potential outcomes and initial level of untreated potential outcomes is the same for the treated group and untreated group.\footnote{We note that other approaches could be used for identification here; for example, one could assume that treatment is as good as randomly assigned after conditioning on covariates and lagged outcomes or one could assume a conditional version of the Change in Changes model (\citet{athey-imbens-2006}) as in \citet{melly-santangelo-2015}.}  What these assumptions mean in the context of job displacement is that (i) the path of earnings that displaced workers would have experienced had they not been displaced is the same as the path of earnings that non-displaced workers with the same characteristics did experience, and (ii) if, for example, the largest increases in earnings for non-displaced workers tended to go to individuals with the highest earnings in the previous period, the same would have occurred for displaced workers if they had not been displaced from their job.

One challenge is that we cannot implement the estimation procedure proposed in \citet{callaway-li-oka-2018} because their approach is for the case with discrete covariates only and involves sample splitting.  This is not feasible in the current application where there are some continuous covariates (e.g., age) and too many discrete covariates relative to the number of observations to split the sample on every possible combination of the discrete covariates.  Instead, we use quantile regression (\citet{koenker-bassett-1978,koenker-2005}) to estimate each required conditional distribution / quantile function.  Quantile regression provides a flexible way to model the conditional distributions and quantiles that we need to estimate in the first step.\footnote{Other papers that have used quantile regression as a first-step estimator include \citet{melly-santangelo-2015,wuthrich-2019}.}

After the conditional distributions of treated and untreated potential outcomes have been obtained in the first step, we construct the joint distribution of earnings following job displacement and earnings that an individual would have had if they had not been displaced and maintained their rank in the earnings distribution over time.  In particular, we impute an individual's counterfactual earnings  by taking their (conditional) rank in the first time period and evaluating the conditional quantile of counterfactual outcomes (which is identified) at that particular rank.\footnote{See also \citet{melly-santangelo-2014} for an example of imputing missing wages using Change in Changes.}  This step results in observed post-displacement earnings being matched with an estimate of earnings in the absence of displacement.  Once this step has been completed, we can obtain any feature of this distribution.  This also implies that the distribution of the effect of job displacement conditional on covariates is identified as well.  In a final step, we again use quantile regression to summarize how the effect of job displacement varies with covariates at different parts of the conditional distribution.

We use the Displaced Workers Survey (DWS) which is a supplement to the Current Population Survey (CPS) to estimate the entire distribution of the effect of job displacement on weekly earnings for displaced workers and evaluate how the effects of job displacement vary across individuals with different observable characteristics (e.g., age, sex, race, education, and marital status) and as a function of what a displaced worker's earnings would have been in the absence of job displacement.  Our main results use the 2016 DWS.  We find that, on average, displaced workers lose about \$157 per week relative to what they would have earned if they had not been displaced (this corresponds to 18\% of pre-displacement earnings).  We also find that there is substantial heterogeneity.  We estimate that 42\% of displaced workers have \textit{higher} earnings following displacement than they would have had if they had not been displaced and maintained their rank in the earnings distribution over time.  On the other hand, we also estimate that 21\% of workers earn at least \$500 less per week than they would have earned if they had not been displaced and had maintained their rank.  These results suggest that the effect of job displacement can be much more severe than the average effect for some workers who are the most negatively affected while being quite mild for other workers.

Since our results on heterogeneous effects of job displacement come from comparing observed earnings of displaced workers to what their earnings would have been if they had not been displaced and had maintained their rank in the earnings distribution from the previous period, it is possible that some of our results are driven by individuals changing their ranks in the earnings distribution over time rather than being due to heterogeneous effects of job displacement.  We propose several robustness checks in order to assess if there is a meaningful amount of heterogeneity apart from the rank invariance conditions.  For one, we compare how much non-displaced workers change their ranks in the earnings distribution over time relative to displaced workers.  We document stronger positive dependence in earnings over time among non-displaced workers (though less than rank invariance) relative to displaced workers.  For another, we compute analogous measures of treatment effect heterogeneity but for non-displaced workers under rank invariance relative to their observed earnings.  Under rank invariance for non-displaced workers, their would be no treatment effect heterogeneity in this case.  Instead, we observe some treatment effect heterogeneity for non-displaced workers, but the amount of treatment effect heterogeneity is noticeably less than for displaced workers.  Both of these robustness checks suggest that violations of rank invariance may partially (but not fully) explain the heterogeneous effects that we document.

In a final step, we use quantile regression to study the effect of covariates on the distribution of the individual-level effect of job displacement.  We find substantial heterogeneity across observed characteristics.  The largest effects of job displacement appear to be concentrated among older, male, college graduates as well as those who would have had high earnings if they had not been displaced.  As an additional check, we compare our results to those obtained under the condition that a worker who is displaced from their job would have earnings that are exactly equal to their pre-displacement earnings.\footnote{An alternative way to think about this is asking the question:  What is an individual's earnings following job displacement relative to what they earned before they were displaced?  This sort of question does not require any identifying assumption (both of these are observed outcomes), but it has the limitation that this difference likely mixes effects of job displacement with trends in earnings over time.}  We also extend our results to different time periods using the 2010 DWS (corresponding to workers displaced during the Great Recession) and the 1998 DWS (corresponding to workers displaced in a substantially earlier time period when the U.S. economy was strong).  We find larger effects of job displacement during the Great Recession, but, interestingly, we continue to find a substantial fraction, 28\%, that appear to have higher earnings following displacement than they would have had if they had not been displaced.

\subsection*{Related Work on Job Displacement}

Our work is related to a large literature on job displacement.  We only briefly survey the most relevant research here.  Almost all work on job displacement considers a single ``effect'' of job displacement.  This effect usually corresponds to an estimated parameter in an OLS regression.  Much of the literature (e.g., \citet{jacobson-lalonde-sullivan-1993}) has been interested in the long-term effect of job displacement.  This is particularly true for papers that have used widely available panel data sets such as the NLSY and PSID (e.g., \citet{ruhm-1991,stevens-1997,kletzer-fairlie-2003}) as well as administrative data (e.g., \citet{wachter-song-manchester-2009,couch-placzek-2010}).  On the other hand, research that has used the Displaced Workers Survey has focused on short run effects of job displacement because the DWS does not contain follow up interviews with displaced workers.  Thus, here, we consider only short-run heterogeneous effects of job displacement.  It would be interesting to extend our approach to the long-term case though this would require an alternative data source.

Two strands of the literature are most related to the current paper.  First, in a series of work (\citet{farber-1997,farber-2005,farber-2017}), Henry Farber has used the DWS to study the effect of job displacement.  Broadly, the main concerns of these papers are the employment and earnings effects of job displacement and how these have varied over time.  \citet{farber-2017} also considers the fraction of displaced workers whose earnings are higher following job displacement than they were in their pre-displacement job.  A primary difference between our main results and those in \citet{farber-2017} is that we try to adjust for how earnings for displaced workers would have evolved over time in the absence of job displacement.  Our results that also use raw changes in earnings over time are also somewhat different; in particular, \citet[Section 7]{farber-2017} focuses mostly on displaced workers that have moved back into full time work while we focus on all displaced workers that are still in the labor force.  In addition, we think about the entire distribution of the effect of job displacement and how the distribution of the effect of job displacement varies across covariates and varies for individuals who would have had high or low earnings if they had not been displaced from their job.  On the other hand, \citet{farber-2017} carefully considers the role of measurement error and shows that this may somewhat reduce the fraction of displaced workers who actually have higher earnings following job displacement than they had in their pre-displacement job.  Second, the application in \citet{callaway-2020} also considers heterogeneous effects of job displacement.  The assumptions in that paper are weaker than in the current paper (particularly in the case when we invoke the assumption of rank invariance over time) though his approach leads to bounds on heterogeneous treatment effect parameters rather than point identification which implies that we are able to learn more about the heterogeneous effect of job displacement.  In addition, using our approach, we are also able to relate heterogeneous effects to covariates as well as to what earnings would have been in the absence of job displacement.

\section{Identification}

In this section, we discuss the parameters that we are interested in identifying and the assumptions required to identify them.  But first, we introduce some notation.  We consider the case where a researcher has access to two periods of panel data which corresponds to the available data that we use from the Displaced Workers Survey.  We denote the two periods by $t$ and $t-1$.  We define outcomes in each period by $Y_{it}$ and $Y_{it-1}$.  Individuals have treated and untreated potential outcomes, $Y_{is}(1)$ and $Y_{is}(0)$, for $s \in \{t,t-1\}$; these correspond to the outcomes that a particular individual would experience if they were displaced or not displaced at a particular point in time.  We consider the case where no one is treated in the first period and some individuals become treated in the second period.\footnote{This is essentially the setup in the case of our application on job displacement due to the Displaced Workers Survey being administered only every other year.  In addition, part of our identification argument utilizes a Difference in Differences-type identification strategy.  This sort of identification strategy identifies effects for individuals who move from being untreated to treated over time.  It does not identify effects for individuals who are treated in the first period; moreover, this group is not used for identifying counterfactual outcome distributions (see discussion below) and therefore can be dropped from the analysis.}  Let the variable $D_i$ denote whether or not an individual is treated (i.e., displaced).  Thus, we observe $Y_{it} = D_iY_{it}(1) + (1-D_i)Y_{it}(0)$ and $Y_{it-1} = Y_{it-1}(0)$.  Define $\Delta Y_{it} := Y_{it} - Y_{it-1}$ and $\Delta Y_{it}(0) := Y_{it}(0) - Y_{it-1}(0)$.  We also observe some covariates $X_i$ that include a worker's race, sex, education, age, and marital status.  

Our interest centers on features of the distribution of $(Y_t(1) - Y_t(0) | D=1)$.  This is the distribution of the treatment effect conditional on being part of the treated group.  In other words, it is the distribution of the difference between earnings of displaced workers following displacement and what their earnings would have been if they had not been displaced.  However, identifying features of this distribution is challenging -- primarily because, for each individual, either $Y_{it}(1)$ or $Y_{it}(0)$ is observed but $Y_{it}(1) - Y_{it}(0)$ is not observed for any individual.  Parameters that depend on this distribution include many treatment effect heterogeneity-type parameters:  (i) the fraction of displaced workers that have higher earnings following displacement than they would have had if they had not been displaced, and (ii) how much lower earnings are for individuals who are most affected by job displacement, among others.  We are also interested in features of the distribution of $(Y_t(1) - Y_t(0) | Y_t(0), D=1)$.  These sorts of parameters include the distribution of the effect of job displacement for displaced workers as a function of what a displaced worker's earnings would have been if they had not been displaced.  Finally, we are interested in the distribution of the treatment effect for displaced workers as a function of covariates; i.e., features of the distribution of $(Y_t(1) - Y_t(0) | X, D=1)$.  In particular, these parameters allow us to think about what individual-level characteristics are associated with the largest effects of job displacement.  Our approach allows one to look at different parts of the conditional distribution of the effect of job displacement which is substantially more informative than looking at average effects across different values of covariates which is commonly done in applied work (see, for example, the discussion in \citet{bitler-gelbach-hoynes-2017}).

What each of these parameters have in common is that they depend on the joint distribution $(Y_t(1),Y_t(0)|X,D=1)$.  By Sklar's Theorem (\citet{sklar-1959}), joint distributions can be written as the copula (which captures the dependence of two random variables; see \citet{nelsen-2007} for more details about copulas) of the marginal distributions.  Thus, we take a two step approach to identifying this joint distribution.  In the first step, we identify the marginal distributions $F_{Y_t(1)|X,D=1}$ and $F_{Y_t(0)|X,D=1}$.  In the second step, we impose additional assumptions to additionally identify their joint distribution.

\subsection{Step 1:  Identifying Marginal Distributions}

The first step in our analysis is to show that $F_{Y_t(1)|X,D=1}$ and $F_{Y_t(0)|X,D=1}$ are identified.  $F_{Y_t(1)|X,D=1}$ is immediately identified from the sampling process.  It is just the conditional distribution of displaced potential earnings for the group of individuals that are displaced from their job -- these are the earnings that are observed for displaced workers.

On the other hand, identifying $F_{Y_t(0)|X,D=1}$ is substantially more challenging.  This is the distribution of earnings that displaced workers would have experienced if they had not been displaced.  We make the following two assumptions:
\begin{assumption}[Distributional Difference in Differences]\label{ass:ddid}
  \begin{align*}
    \Delta Y_t(0) \independent D | X
  \end{align*}
\end{assumption}
\begin{assumption}[Copula Invariance Assumption]\label{ass:cia}
Let $C_{\Delta Y_t(0), Y_{t-1}(0)|X,D=d}$ denote the copula between the change in untreated potential outcomes and the initial level of untreated potential outcomes conditional on covariates and treatment status.  Then,
  \begin{align*}
    C_{\Delta Y_t(0), Y_{t-1}(0)|X,D=1} = C_{\Delta Y_t(0), Y_{t-1}(0)|X,D=0} \quad a.s.
  \end{align*}
\end{assumption}
\begin{assumption}[Observed Data] \label{ass:iid} The observed data consists of $\{Y_{it}, Y_{it-1},X_i,D_i\}_{i=1}^n$ and are independent and identically distributed and where $n$ denotes the sample size.
\end{assumption}
\begin{assumption}[Continuously Distributed Outcomes] \label{ass:continuous}
  $Y_t(0)$ and $Y_{t-1}(0),$ are continuously distributed conditional on $X$ and $D=d$ for $d\in\{0,1\}$
\end{assumption}

\Cref{ass:ddid,ass:cia,ass:iid,ass:continuous} are the assumptions made in \citet{callaway-li-oka-2018} to identify the counterfactual distribution $F_{Y_t(0)|X,D=1}$.  The first is a Difference in Differences assumption.  It says that the distribution of the path of earnings that individuals who are displaced from their job would have experienced if they had not been displaced from their job is the same as the distribution of the  path of earnings that individuals that were not displaced from their job actually experienced (after conditioning on covariates).  The second assumption is a Copula Invariance assumption.  It says that if, for example, we observe the largest earnings increases over time going to those at the top of the earnings distribution for the group of non-displaced workers, then we would have observed the same thing for displaced workers if they had not been displaced (again, this is conditional on covariates).  \Cref{ass:iid} says that we observe two periods of panel data that is cross sectionally iid.  \Cref{ass:continuous} says that untreated potential outcomes are continuously distributed.  Importantly, in the case of job displacement, \Cref{ass:continuous} does not require that treated potential outcomes (i.e., outcomes following displacement for displaced workers) be continuously distributed.  This allows for some displaced workers to not be working following displacement which leads to a mass of workers with zero earnings.  \citet{callaway-li-oka-2018} show that, under these conditions
\begin{align}\label{eqn:cf}
  F_{Y_t(0)|X,D=1}(y|x) = P( \Delta Y_t + Q_{Y_{t-1}|X,D=1}(F_{Y_{t-1}|X,D=0}(Y_{t-1}|x)|x) \leq y | x, D=0)
\end{align}
where every term in \Cref{eqn:cf} is identified and $Q(\cdot)$ denotes the quantile function (i.e., the inverse of the cdf).   %
In other words, we can recover the distribution of outcomes that displaced workers would have experienced in the absence of displacement.  And this implies that both marginal distributions of interest are identified.  It is worth briefly considering where the expression for this counterfactual distribution comes from.  Omitting conditioning on covariates (for simplicity), notice that the counterfactual distribution $F_{Y_t(0)|D=1}(y)$ is given by
\begin{align*}
  P(Y_t(0) \leq y | D=1) &= P(\Delta Y_t(0) + Y_{t-1}(0) \leq y | D=1) \\
                         &\hspace{-25pt} = P( Q_{\Delta Y_t(0)|D=1}( F_{\Delta Y_t(0)|D=0}(\Delta Y_t(0))) + Q_{Y_{t-1}(0)|D=1}(F_{Y_{t-1}(0)|D=0}(Y_{t-1}(0))) \leq y | D=0) \\
                         &\hspace{-25pt} = P( \Delta Y_t(0) + Q_{Y_{t-1}(0)|D=1}(F_{Y_{t-1}(0)|D=0}(Y_{t-1}(0))) \leq y | D=0)
\end{align*}
The first equality adds and subtracts $Y_{t-1}(0)$.  The second equality is most interesting and holds under \Cref{ass:cia} which says that the joint distribution of $(\Delta Y_t(0), Y_{t-1}(0))$ is the same across displaced and non-displaced individuals after adjusting for differences in the marginal distributions -- adjusting for differences in the marginal distributions is exactly what the two terms involving $Q(F(\cdot))$ do.  Finally, the third equality holds immediately by \Cref{ass:ddid}.   Thus, the result in \Cref{eqn:cf} says that the distribution of outcomes that displaced workers would have experienced in the absence of job displacement can be obtained from the distribution of earnings of non-displaced workers but only after ``adjusting'' for differences in the marginal distributions of earnings in the period before displacement as well as accounting for changes in earnings over time experienced by non-displaced workers.

\subsection{Step 2:  Identifying the Joint Distribution}

Identifying the joint distribution of treated and untreated potential outcomes is a notoriously difficult challenge in the microeconometrics literature.  Standard identifying assumptions (or even random assignment of the treatment) do not identify this joint distribution.  This section describes our approach to identifying this distribution.  Define
\begin{align}\label{eqn:ytild}
  \widetilde{Y}_{it}(0) := Q_{Y_t(0)|X,D=1}(F_{Y_{t-1}|X,D=1}(Y_{it-1}|X_i)|X_i)
\end{align}
Notice that $F_{Y_{t-1}|X,D=1}(Y_{it-1}|X_i)$ is a displaced worker's rank in the conditional earnings distribution in the period before anyone is displaced.  Thus, conditional on our identification arguments in the previous section for the counterfactual distribution, the transformation on the right hand side of \Cref{eqn:ytild} is available and $\widetilde{Y}_{it}(0)$ is the outcome that a displaced worker would experience if they had not been displaced and if they had the same rank in the earnings distribution as they did in the previous period.

Immediately, \Cref{eqn:ytild} implies that the joint distribution $(Y_t(1),\widetilde{Y}_t(0)|X,D=1)$ is identified.  One idea is to just call features of this joint distribution as the parameters of interest.  In other words, without any additional assumptions, we can identify parameters such as the fraction of displaced individuals that have higher earnings than they would have had if they had not been displaced and if they had maintained the same rank in the earnings distribution as they had in the pre-treatment period.

Alternatively, we can make the additional assumption:
\begin{assumption}[Rank Invariance Over Time]\label{ass:riot}
  \begin{align*}
    F_{Y_t(0)|X,D=1}(Y_t(0)|X) = F_{Y_{t-1}(0)|X,D=1}(Y_{t-1}(0)|X)
  \end{align*}
\end{assumption}
\Cref{ass:riot} says that, in the absence of job displacement, displaced workers would have maintained their rank in the distribution of earnings over time.  \Cref{ass:riot} implies that $Y_{it}(0) = \widetilde{Y}_{it}(0)$ and, thus, that the joint distribution $(Y_t(1),Y_t(0)|X,D=1)$ is identified, and we can obtain any features of this joint distribution.  For comparison, identifying this distribution allows one to ask:  What fraction of displaced workers actually benefit from job displacement without the additional caveat mentioned above.  At any rate, employing \Cref{ass:riot} leads to exactly the same results as focusing on features of the joint distribution $(Y_t(1),\widetilde{Y}_t(0)|X,D=1)$ -- only the interpretation is different.  We use the notation $Y_{it}(0)$ rather than $\widetilde{Y}_{it}(0)$ throughout the rest of the paper for simplicity but we interpret our results carefully in the application as the caveat of being conditional on having the same rank over time likely applies.

\begin{remark}
  An alternative approach is to define $\widetilde{Y}_{it}(0) = Y_{it-1}(0)$ among displaced workers.\footnote{\citet[Section 7]{farber-2017} considers the fraction of displaced workers who have higher earnings following displacement than they did in their previous job which is similar to what we propose here.}   In this case, one can immediately identify the joint distribution $(Y_t(1),\widetilde{Y}_t(0)|X,D=1)$.  Relative to the previous case, there are some advantages and disadvantages to this approach.  The main advantage is that it does not require \Cref{ass:ddid,ass:cia} as the distribution of $(Y_t(1), \widetilde{Y}_t(0)|X,D=1)$ is identified from the sampling process rather than requiring identification assumptions.  Without further assumptions, one can immediately identify things like: the fraction of displaced workers who have higher earnings following job displacement than they had before they were displaced, etc.  The main disadvantage is that, in this case, $Y_{it}(0)=\widetilde{Y}_{it}(0)$ seems less plausible.  But, at any rate, this approach serves as a useful comparison.
\end{remark}

\begin{remark} \Cref{ass:riot} is distinct from the more commonly invoked ``cross-sectional'' rank invariance assumption -- namely, that a displaced worker's rank in the distribution of $Y_{it}(1)$ (which is observed) is the same as their rank would be in the distribution of $Y_{it}(0)$ if they had not been displaced.  This is a very strong (and implausible) assumption in the current context as it rules out things like individuals not finding work following displacement.
\end{remark}

\subsection{Comparison to Quantile Treatment Effects}

A common alternative strategy for trying to analyze heterogeneous effects of experiencing some treatment is to compare the distribution of treated potential outcomes to the distribution of untreated potential outcomes; in our case, this would mean comparing the distribution of earnings that displaced workers actually experienced to the distribution of earnings that they would have experienced in the absence of job displacement.\footnote{Identifying these distributions typically either requires experimental data or some identifying argument.  However, there are many approaches that are available in the econometrics literature for identifying these distributions; e.g., \citet{firpo-2007} under selection on observables, \citet{abadie-2003,frolich-melly-2013,wuthrich-2019} when a researcher has access to an instrument, \citet{athey-imbens-2006,bonhomme-sauder-2011,melly-santangelo-2015,callaway-li-oka-2018,callaway-li-2019} when a researcher has access to more than one period of data.  The latter set of papers is most relevant in the current application.}  Unlike our main parameters of interest, comparing these two distributions does not require identifying the joint distribution of treated and untreated potential outcomes -- in fact, these are the two distributions that are identified using only Step 1 of our identification argument above.  Often researchers invert these two distributions and take their difference.  The result of this transformation is a class of parameters called quantile treatment effects which are important parameters in the program evaluation literature.  

In the context of job displacement, these types of parameters may have severe limitations.  These limitations stem from the fact that individuals may have substantial differences in their ``rank'' if they were displaced relative to if they were not displaced.\footnote{This sort of issue is well-known in the econometrics literature (e.g., \citet{heckman-smith-clements-1997}).  In other cases, comparing these two distributions may be very useful for policy evaluation or to perform social welfare calculations (\citet{sen-1997,carneiro-hansen-heckman-2001}).  However, this does not apply for job displacement primarily because job displacement is not a ``policy'' (if it were, our results would indicate that it is a really bad policy).  Instead, the reason why researchers would be interested in distributions here is to examine heterogeneous effects as we do in the current paper, and simply comparing the marginal distributions is of limited usefulness in this case.}  This means that, although it is tempting to interpret differences between, for example, the 10th percentile of displaced potential earnings and non-displaced potential earnings as the effect of job displacement for individuals at the lower part of the earnings distribution, this sort of interpretation is unlikely to be correct.  The reason is that some individuals in the lower part of the earnings distribution might have had relatively high earnings if they had not been displaced.  When some individuals remain unemployed or have substantially reduced hours following job displacement, this can systematically lead to larger differences between the observed distribution and counterfactual distribution occurring in the lower part of the distribution irrespective of whether the effect of job displacement is bigger or smaller for low or high earnings workers.  Instead, to think about these sorts of effects, we propose to estimate the distribution of the individual-level effect of job displacement for displaced workers conditional on what earnings would have been in the absence of displacement.  Because of our above results on identifying the joint distribution of treated and untreated potential outcomes, these sorts of parameters are identified and estimable in our framework.

\section{Estimation}

Given the identification results above, determining the best estimation strategy is still challenging.  Like many applications in economics, we have a relatively large number of covariates that we would like to condition on to make the identifying assumptions more plausible and only a moderate number of observations.  The first step is to estimate the counterfactual distribution of untreated potential outcomes conditional on covariates for the treated group as in \Cref{eqn:cf}.  This requires estimating three conditional distribution/quantile functions.  These are challenging.  Distributional assumptions about the distribution of outcomes conditional on covariates are likely to be invalid.  On the other hand, implementing fully nonparametric estimators is infeasible in our case.  \citet{callaway-li-oka-2018} consider the case with only discrete covariates and propose nonparametric estimators based on splitting the sample according to each possible combination of discrete covariates.  This is infeasible in our case because we have some continuous covariates and the number of discrete covariates is too large relative to the number of observations to split the sample.  Instead, we propose to estimate these conditional distributions/quantiles using quantile regression.  In particular, we suppose that
\begin{assumption}[Quantile Regression]\label{ass:qr}
  For all $(s,d) \in \{t,t-1\} \times \{0,1\}$,
  \begin{align*}
    Y_{is}(0) = Q_{Y_s(0)|X,D=d}(U_{isd}|X_i) = P_{sd}(X_i)'\beta_{sd}(U_{isd})
  \end{align*}
  with $U_{isd}|X_i,D_i \sim U[0,1]$ and where $P_{sd}(X)$ are some transformations of the vector of covariates $X$.  
\end{assumption}
\Cref{ass:qr} suggests that we can estimate each of the terms in \Cref{eqn:cf} using quantile regression.  In particular, consider the following algorithm to estimate the counterfactual distribution in \Cref{eqn:cf} and estimate $Y_{it}(0)$ for each individual in our sample.
\begin{algorithm}\label{alg:1} Let $\tau$ denote a fine grid of equally spaced elements between 0 and 1,
  \begin{enumerate}
  \item Estimate $\hat{Q}_{Y_{t-1}|X,D=1}(\tau)$ by quantile regression of $Y_{it-1}$ on $X_i$ using displaced individuals
  \item Estimate $\hat{Q}_{Y_{t-1}|X,D=0}(\tau)$ by quantile regression of $Y_{it-1}$ on $X_i$ using non-displaced individuals and invert to obtain $\hat{F}_{Y_{t-1}|X,D=0}$.
  \item For non-displaced individuals, set $\hat{\widetilde{Y}}_{it-1} := \hat{Q}_{Y_{t-1}|X,D=1}(\hat{F}_{Y_{t-1}|X,D=0}(Y_{it-1}|X_i)|X_i)$
  \item Estimate $\hat{Q}_{Y_{t}(0)|X,D=1}(\tau)$ by quantile regression of $\Delta Y_{it} + \hat{\widetilde{Y}}_{it-1}$ on $X_i$ using non-displaced individuals
  \item For displaced individuals, set $\hat{Y}_{it}(0) := \hat{Q}_{Y_{t}(0)|X,D=1}(\hat{F}_{Y_{t-1}|X,D=1}(Y_{it-1}|X_i)|X_i)$
  \end{enumerate}
\end{algorithm}
\Cref{alg:1} estimates $Y_{it}(0)$ for all individuals in the treated group.  Thus, we can immediately obtain features of the unconditional distribution of $(Y_t(1),Y_{t}(0)|D=1)$ using the draws of $(Y_{it},\hat{Y}_{it}(0))$ for displaced workers.  For example, we can estimate the fraction of individuals that have higher earnings following job displacement by
\begin{align*}
  \frac{1}{n_1}\sum_{i \in \mathcal{D}} \indicator{Y_{it} - \hat{Y}_{it}(0) > 0}
\end{align*}
where $n_1$ denotes the number of displaced workers and $\mathcal{D}$ denotes the set of displaced workers.  We are also interested in how features of the distribution of $Y_{it}(1) - Y_{it}(0)$ vary with covariates.  It is straightforward to run OLS regressions of $Y_{it}(1) - \hat{Y}_{it}(0)$ on covariates.  These results are interesting, and we report them in the application.  However, our identification results allow us to go substantially further than this.  In particular, we can also estimate how covariates affect the entire distribution of the effect of job displacement.  To estimate this, we again can use quantile regression.  In particular, we suppose that
\begin{align*}
  Q_{(Y_t(1) - Y_t(0))|X,D=1}(\tau|x) = P_{\Delta_X}(x)'\beta_{\Delta_X}(\tau)
\end{align*}
where $P_{\Delta_X}$ denotes transformations of the covariates.  The estimates of these conditional quantiles and quantile regression parameters are of interest.  For example, if we set $\tau$ to be small (e.g., 0.05), $\beta_{\Delta_X}(\tau)$ provides the effect of covariates on the individual-level effect of job displacement for individuals who are most negatively affected by job displacement.

Finally, we are interested in features of the distribution of $Y_{it}(1) - Y_{it}(0)$ across different values of $Y_{it}(0)$; that is, how the effect of job displacement changes across different values of what a displaced worker's earnings would have been if they had not been displaced.  Once again, one can estimate OLS regressions of $Y_{it}(1) - \hat{Y}_{it}(0)$ on $\hat{Y}_{it}(0)$; however, we also use quantile regression to study how the entire distribution of the effect of job displacement is related to non-displaced potential earnings.  Here we suppose that
\begin{align*}
  Q_{(Y_t(1) - Y_t(0))|Y_t(0),D=1}(\tau|y) = P_{\Delta_{Y}}(y)'\beta_{\Delta_{Y}}(\tau)
\end{align*}
where $P_{\Delta Y}(y)$ are transformations of non-displaced potential earnings, and $\beta_{\Delta_Y}$ gives the effect of non-displaced potential earnings on particular quantiles of the individual-level effect of job displacement.

In order to conduct inference, we use the bootstrap.  For each of the parameters that we consider, we re-sample the original data a large number of times and, at each iteration, re-estimate that parameter using the full estimation procedure.  This approach accounts for estimation uncertainty coming from both the first and second steps of our estimation procedure.

\section{Data}
We use data from the Displaced Workers Survey (DWS).  The DWS has been administrated every two years since 1984 as a supplement to the Current Population Survey (CPS)  (\citet{flood-king-ruggles-warren-2017})
and contains information on earnings and employment status for displaced workers. Respondents are asked whether they lost their jobs at any point in the last three years due to: (i) plant closing or moving, (ii) insufficient work, or (iii) position being abolished.  These are sometimes referred to as the ``big three'' reasons for job displacement (\citet{farber-2017}).   Our main results below focus on 2015-2016.  We also provide similar results for 2009-2010 (during the Great Recession) and 1997-1998 (an earlier period when the U.S. economy was quite strong).

Although the DWS provides useful information on job displacement status and the current weekly earnings of displaced workers and their weekly earnings on their lost job, it does not provide any information about the earnings of non-displaced workers (though non-displaced workers can be identified).  After identifying non-displaced workers using the DWS, we obtain their earnings by matching the DWS to CPS outgoing rotation group data using individual-level identification numbers which are available from IPUMS.\footnote{Households in the CPS are usually interviewed for four consecutive months, are out of the sample for the next eight months, and then interviewed for another four months. After the eighth time participating, they leave the sample. Households that are interviewed in the fourth or eighth month (which are called outgoing rotation groups) are asked additional labor market questions.  Thus, using our approach, we observe earnings for the untreated group in two consecutive years, but they can be different months for different individuals.} This matching process helps us to find information on the current weekly earnings as well as weekly earnings in the previous year for non-displaced workers.  Effectively, this turns our untreated group into a two period panel dataset.  We also drop all non-displaced workers whose earnings are missing in either period. 
For workers that are displaced, we drop those that are not currently in the labor force and assign 0 earnings to those in the labor force but who are not currently employed.  The timing for available earnings is also somewhat different for displaced and non-displaced workers.  For displaced workers, the DWS asks for their earnings in their pre-displacement job.  But displacement could have occurred any time in the last three years.  On the other hand, for non-displaced workers, we only have their current weekly earnings and their earnings one year ago.  %

{ \onehalfspacing 
\newcolumntype{.}{D{.}{.}{-1}}
\ctable[caption={Summary Statistics},label=table:ss,pos=!tbp,]{lrrrr}{\tnote[]{\textit{Sources: CPS and Displaced Workers Survey}}}{\FL
\multicolumn{1}{l}{}&\multicolumn{1}{c}{Displaced}&\multicolumn{1}{c}{Non-Displaced}&\multicolumn{1}{c}{Difference}&\multicolumn{1}{c}{P-val on Difference}\ML
{\bfseries Earnings}&&&&\NN
~~2016 Earnings&687.82&1000.43&-312.604&0.00\NN
~~2015 Earnings&882.56&956.43&-73.875&0.00\NN
~~Change Earnings&-194.74&43.99&-238.729&0.00\ML
{\bfseries Covariates}&&&&\NN
~~Male&0.57&0.51&0.057&0.00\NN
~~White&0.81&0.82&-0.008&0.48\NN
~~Married&0.51&0.63&-0.113&0.00\NN
~~College&0.33&0.40&-0.066&0.00\NN
~~Less HS&0.07&0.06&0.009&0.21\NN
~~Age&42.11&45.4&-3.294&0.00\NN
~~N&1633&3915&&\LL
}
}

The summary statistics for key
variables of displaced and non-displaced workers in 2016 are presented in \Cref{table:ss}. %
Weekly earnings are about \$74 dollars higher for non-displaced workers than displaced workers in the period before they are displaced.  Following job displacement, this gap in average weekly earnings increases to \$313.  This occurs due to a combination of average weekly earnings increasing by  \$44 for non-displaced workers and decreasing by \$195 for displaced workers (on top of the existing gap in the previous period).  For covariates, we include an individual's sex, race, marital status, age, and education.  Displaced workers are more likely to be male, not married, younger, and not have a college degree; both groups are roughly equally likely to be white or have less than a high school education.

\section{Results}

\begin{figure}[t] 
  \caption{Distributions of Earnings} \label{fig:1}
  \includegraphics[width=0.48\textwidth]{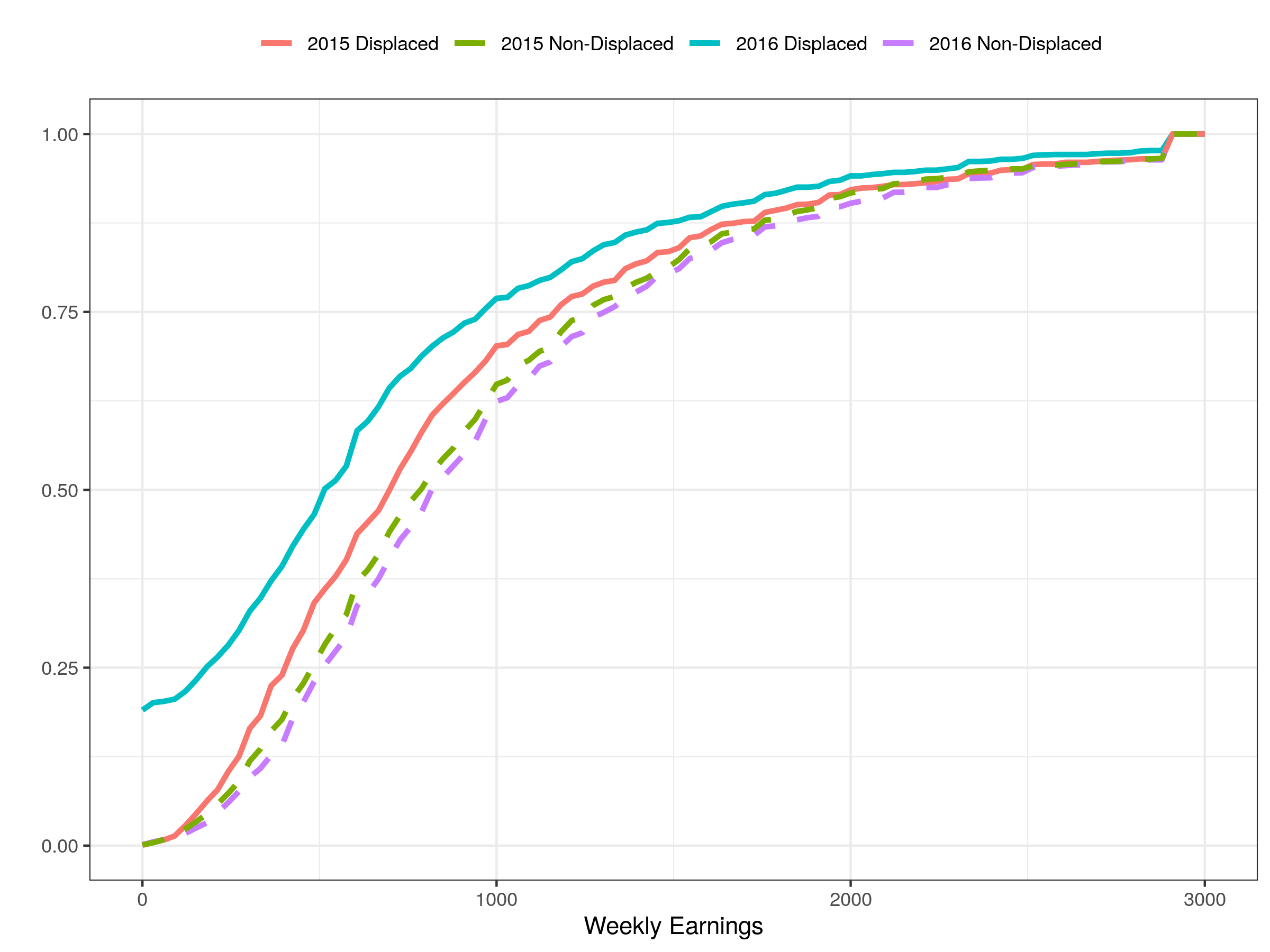}
  \includegraphics[width=0.48\textwidth]{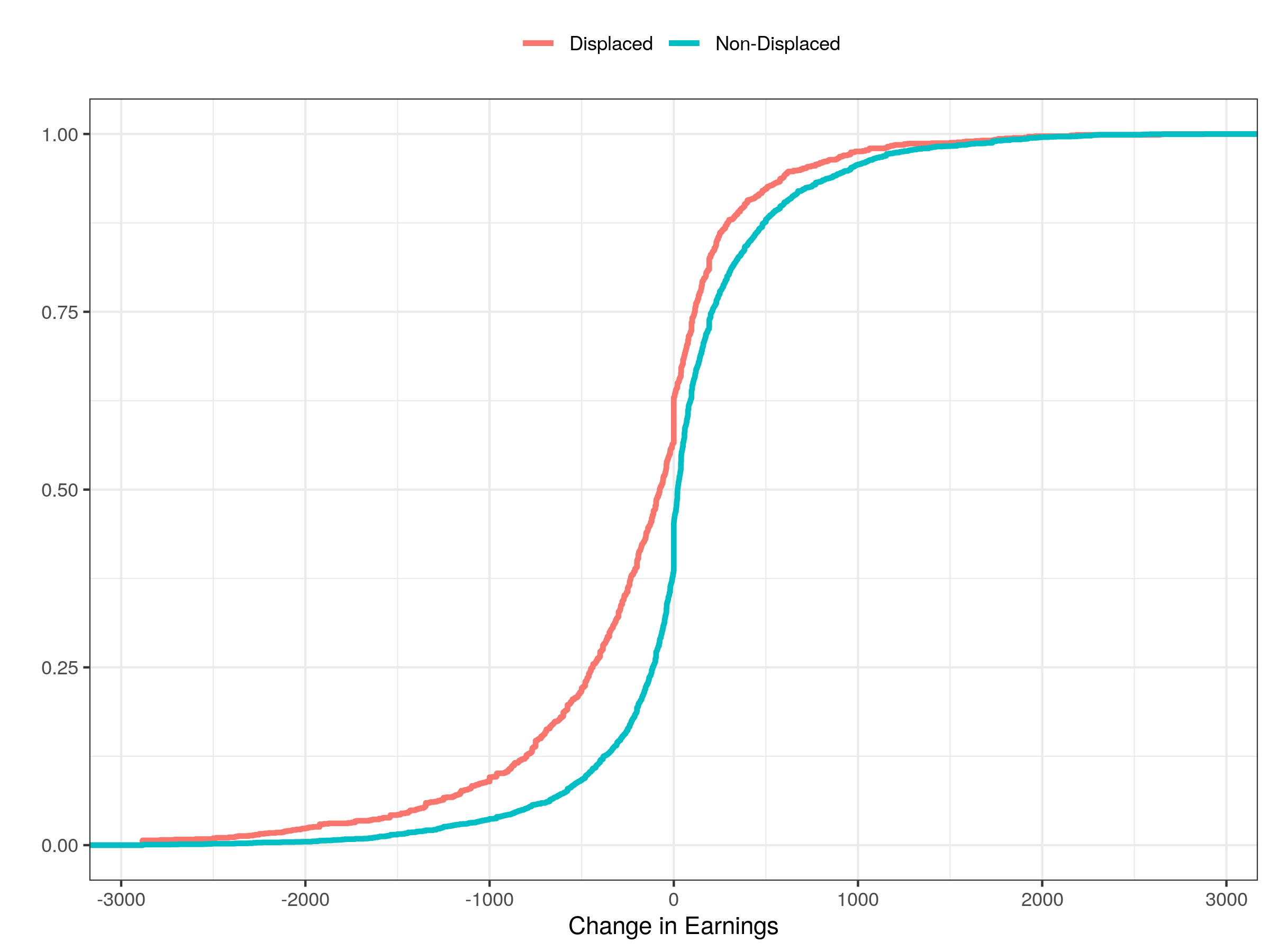}
  \subcaption*{\textit{Notes:}  The left panel contains plots of the distributions of earnings in 2015 and 2016 for displaced and non-displaced workers.  The right panel plots the distribution of the change in weekly earnings over time separately for displaced and non-displaced workers.}
\end{figure}

As a first step, \Cref{fig:1} plots the distributions of weekly earnings for displaced workers pre- and post-displacement.  Clearly, the distribution of post-displacement earnings is shifted to the left.  Notably, there is also a mass point in the distribution of post-displacement earnings -- about 20\% of displaced workers have 0 earnings.  These distributions also exhibit a pattern that is consistent with very heterogeneous effects -- namely, the difference between the distributions is largest in the lower part of the earnings distributions.  This panel also plots the distributions of earnings for non-displaced workers in both periods.  In the first period, non-displaced workers earn more than displaced workers.  And, unlike displaced workers, in the second time period, the distribution of earnings shifts somewhat to the right (indicating somewhat higher earnings).

The right panel of \Cref{fig:1} plots the distribution of the change in weekly earnings over time for displaced workers.  Recall that this distribution does not require any identifying assumption; it is just the difference in earnings over time.  Interestingly, our estimates in the figure indicate that about 37\% of displaced workers have higher weekly earnings than they had before they were displaced.  This number may overstate how many workers actually have higher earnings than they \textit{would have had} if they had not been displaced especially in the case where earnings would have increased in the absence of job displacement.  On the other hand, around 22\% of workers earn at least \$500 less per week than before displacement and about 10\% earn at least \$1000 less per week than before displacement -- these are substantial differences.  Thus, it is immediately clear that the difference in the average decrease in earnings for displaced workers relative to non-displaced workers of \$239 per week (see \Cref{table:ss}) masks substantial heterogeneity.  The right panel also displays the distribution of the change in earnings over time for non-displaced workers.  This distribution is notably different with non-displaced workers being much less likely to experience large declines in their earnings over time and more likely to experience an increase in earnings over time.

\begin{figure}[t] 
  \caption{Counterfactual Distribution of Earnings for Displaced Workers} \label{fig:2}
  \centering \includegraphics[width=0.8\textwidth]{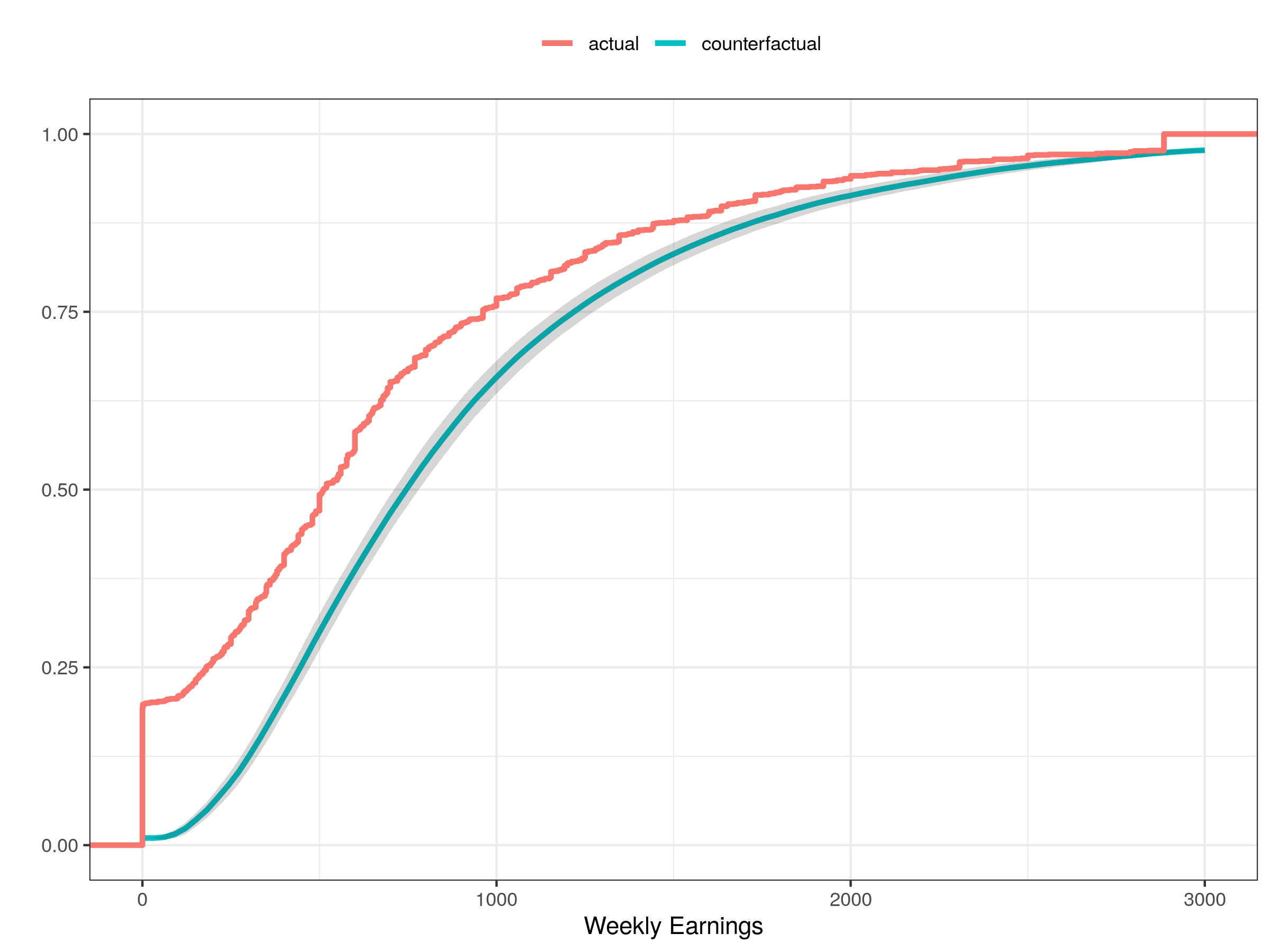}
  \subcaption*{\textit{Notes:}  Plots of the actual and counterfactual distribution of earnings for displaced workers.  The counterfactual distribution also includes a 95\% pointwise confidence interval computed using the empirical bootstrap with 1000 iterations.}
\end{figure}

The next step in our analysis is to construct the counterfactual distribution of weekly earnings that displaced workers would have experienced if they had not been displaced from their jobs.  We do this using the approach of \citet{callaway-li-oka-2018} as described in \Cref{eqn:cf} and \Cref{alg:1}.  \Cref{fig:2} plots this counterfactual distribution.\footnote{The distribution depends on the value of the covariates; the figure reports the average of the distribution over all values of covariates for displaced workers in the dataset.  This is a standard way of converting conditional effects into unconditional effects.}  Notice that the counterfactual distribution sits to the right of the actual distribution of earnings.  Also, the biggest difference between the two distributions occurs in the lower part of the distribution.  Once the counterfactual distribution is identified, this also implies that the average effect of job displacement is identified.  We estimate that job displacement reduces earnings by \$157 per week on average, which is about 18\% of pre-displacement earnings.

\begin{remark} \label{rem:3}
The results in \Cref{fig:2} are also closely related to quantile treatment effects.  The Quantile Treatment Effect on the Treated (QTT) is the difference between the quantiles of observed outcomes for displaced workers and the quantiles of the counterfactual distribution of outcomes that they would have experienced if they had not been displaced.  The QTT comes from inverting the distributions in \Cref{fig:2}.  Because the difference between these two distributions is largest at the lower part of the distribution, it is tempting to interpret this as saying that the effect of job displacement is largest for workers in the lower part of the earnings distribution.  This is an incorrect interpretation though because it ignores the possibility that workers change their ranks in the distribution of earnings due to job displacement (e.g., workers who would have had high earnings if they had not been displaced remaining unemployed following displacement).  In fact, we find evidence (discussed below) that points in exactly the opposite direction -- that the effect of job displacement appears to be larger for high earnings workers than low earnings workers. 
\end{remark}

\begin{figure}[t!] 
  \caption{Distribution of the Effect of Job Displacement}
  \label{fig:dte}
  \centering \includegraphics[width=0.8\textwidth]{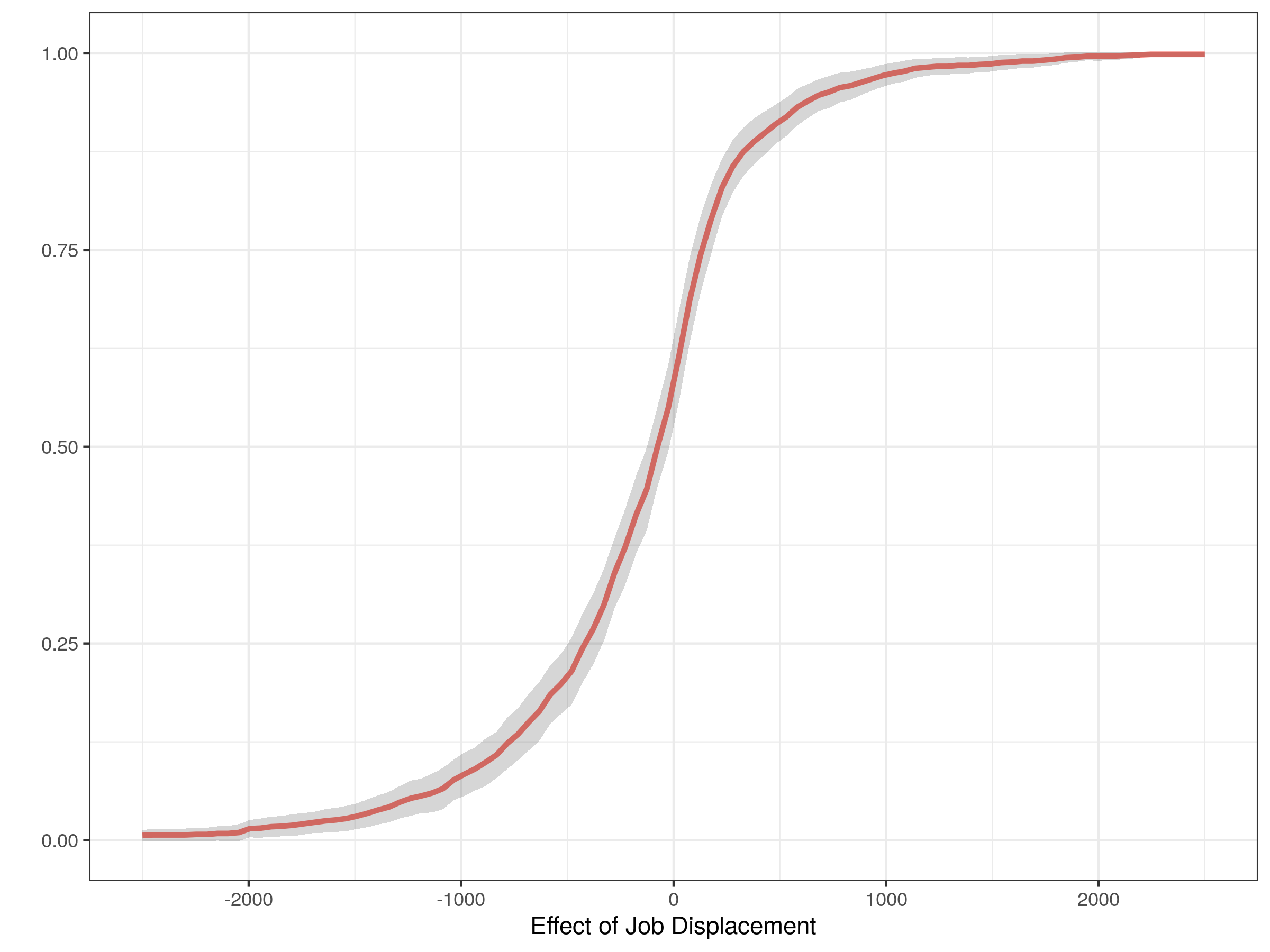}
  \subcaption*{\textit{Notes:}  The red line is the estimated distribution of the effect of job displacement for displaced workers as discussed in the text.  The shaded gray area provides a uniform 95\% confidence band computed using the empirical bootstrap with 1000 observations.}
\end{figure}

Next, we can map each value of $Y_{it}(1)$ to $\hat{Y}_{it}(0)$ and create a pair of $(Y_{it}(1),\hat{Y}_{it}(0))$ for each displaced worker.  \Cref{fig:dte} plots the distribution of $Y_{it}(1) - \hat{Y}_{it}(0)$ for displaced workers.  This is the distribution of the difference between earnings for displaced workers relative to what they would have earned if they had not been displaced and maintained the same rank in the earnings distribution over time.  Under the additional assumption of rank invariance over time, this is the distribution of the effect of job displacement.  Interestingly, this distribution is remarkably close to the distribution of the change in outcomes over time (as in the right panel of \Cref{fig:1}).\footnote{It is worth explaining more why the distributions are so similar.  The change in earnings over time is given $Y_{it} - Y_{it-1}$ for displaced workers.  The estimates in \Cref{fig:dte} come from the distribution of $Y_{it} - \hat{Y}_{it}(0)$ for displaced workers.  Thus, any differences between the two distributions comes from differences between $Y_{it-1}$ and $Y_{it}(0)$.  From \Cref{eqn:ytild}, we generate $Y_{it}(0) = Q_{Y_t(0)|X,D=1}(F_{Y_{t-1}|X,D=1}(Y_{it-1}|X_i)|X_i)$; in practice, this amounts to finding an individuals rank in the conditional distribution of outcomes in time period $t-1$ and applying the conditional quantile function for untreated potential outcomes to that rank.  Also, notice that $Y_{it-1} = Q_{Y_{t-1}|X,D=1}(F_{Y_{t-1}|X,D=1}(Y_{it-1}|X_i)|X_i)$; in other words, an alternative way to obtain an individual's outcome in period $t-1$ is to find their rank in the conditional distribution in time period $t-1$ and apply the conditional quantile for outcomes in period $t-1$ to that rank.  Thus, any difference between the distribution of $Y_{it} - Y_{it-1}$ and $Y_{it} - Y_{it}(0)$ for displaced workers comes down to differences in the conditional quantiles (or distributions) of $Y_{it}(0)$ and $Y_{it-1}$.  In our case, these two distributions are very similar -- suggestive evidence of this can be found in \Cref{fig:1,fig:2} -- and this explains the reason for the high degree of similarity here.}  This implies that our results for the distribution of the actual effect of job displacement are very similar to our earlier estimates that instead use the change in earnings over time for displaced workers.  Here, we estimate that (i) 42\% of displaced workers have higher earnings following displacement than they would have had if they had not been displaced and maintained the same rank in the earnings distribution over time,\footnote{It is perhaps surprising that such a large fraction of displaced workers appear to have higher earnings following displacement.  Comparing pre- and post-displacement earnings of displaced workers, \citet{farber-2017} similarly finds large percentages of workers earning more in their post-displacement job than in their pre-displacement job.  Some possible explanations for this finding that he offers (and that likely apply to our case as well) are:  (i) some displaced workers may have higher earnings but lower utility in their new job, (ii) search costs may keep workers from looking for new jobs that could potentially offer higher earnings, and (iii) risk aversion in the sense that workers have more information about their current job and may be reluctant to leave it for the possibility of a higher paying job when they have less information about other aspects of the new job.  Another possible explanation is measurement error in earnings though \citet{farber-2017} considers a sensitivity analysis related to measurement error and finds only fairly moderate changes to these results even under conservative assumptions on measurement error in earnings. } (ii) 21\% have at least \$500 less in weekly earnings than they would have had if they had not been displaced and maintained the same rank in the earnings distribution over time, and (iii) 8\% have at least \$1000 less weekly earnings than they would have had if they had not been displaced and maintained the same rank in the earnings distribution over time.

Taken together, our results so far indicate that there is substantial heterogeneity in the effect of job displacement.  Even though there is a fairly large average effect of job displacement, a substantial fraction of workers appear to be no worse off due to job displacement than they would have been if they had not been displaced.  On the other hand, the negative effects of job displacement seem to be largely concentrated among 10-20\% of displaced workers who experience large negative effects.

\begin{figure}[t!]
  \caption{Quantile Regression Estimates of Job Displacement Effects on Covariates} \label{fig:4}
  \centering \includegraphics[width=\textwidth]{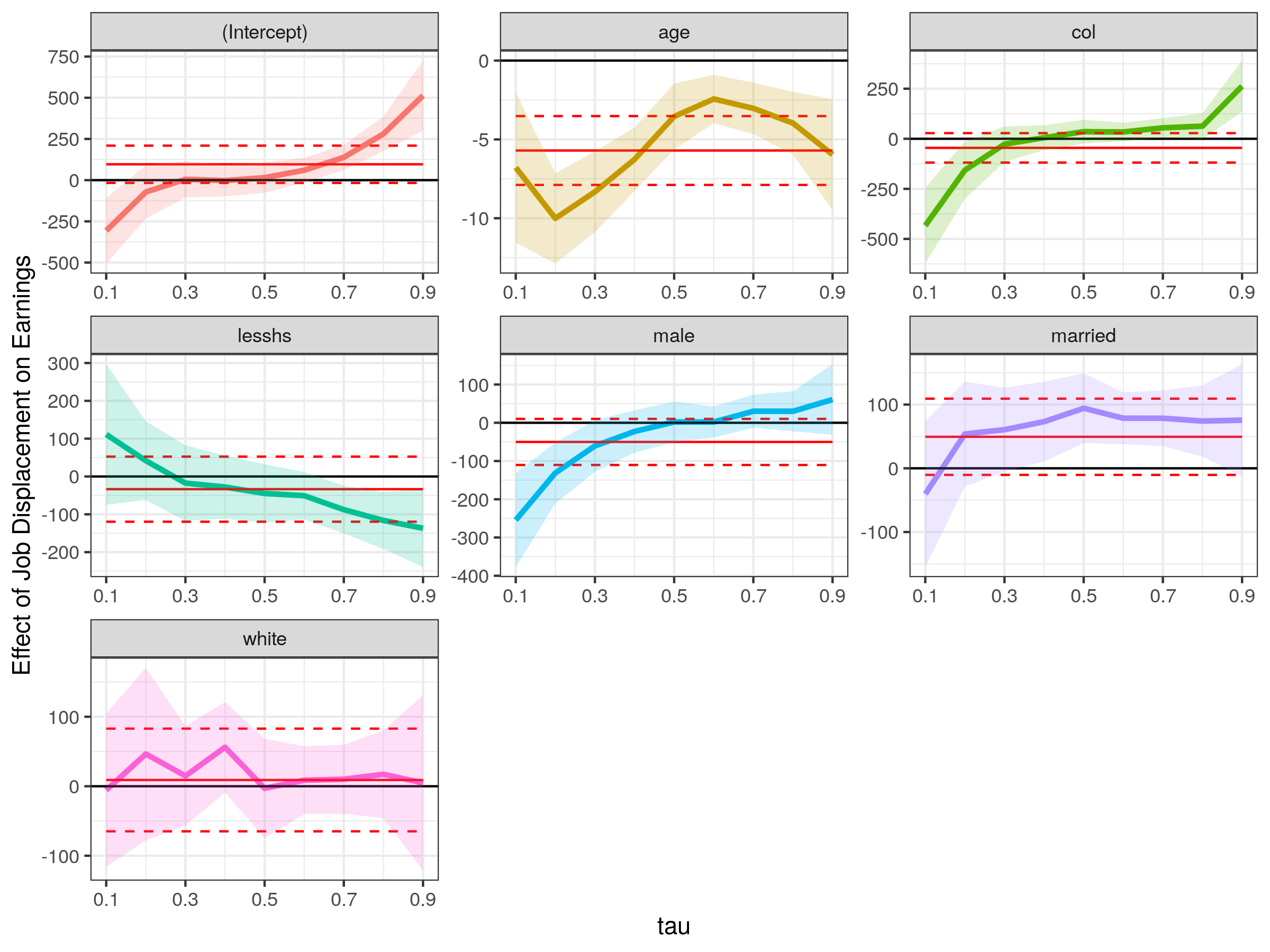}
  \subcaption*{\textit{Notes:} QR estimates of the effect of covariates on the quantiles of the effect of job displacement.  The solid red horizontal lines provide OLS estimates of the effect of each covariate, and the horizontal dashed line contains a 90\% confidence interval.  The other multi-colored lines provide QR estimates of the effect of each covariate at particular quantiles from 0.1, 0.2, \ldots, 0.9.  The shaded area contains pointwise 90\% confidence intervals from the quantile regressions.}
\end{figure}

Next, we use quantile regression to see how covariates affect the distribution of the difference between earnings following displacement for displaced workers and what they would have been in the absence of displacement and if workers had maintained their rank in the earnings distribution over time.  \Cref{fig:4} provides estimates of the effect of each covariate across different quantiles of this conditional distribution.  There are some quite interesting patterns here.  On average, the effect of job displacement is somewhat more negative for men than for women, but the effect is very heterogeneous.  For individuals that experience large negative effects, the effects for men are much more negative than the effects for women.\footnote{Recall that the outcome in these quantile regressions is $Y_{it}(1) - \hat{Y}_{it}(0)$ for displaced individuals.  This means that ``large negative effects'' occur at the lower quantiles in \Cref{fig:4}.}  On the other hand, for individuals that are towards the top of the distribution of the treatment effect (i.e., experience small or positive effects of job displacement), men tend to have larger benefits than women.  Differences in the effect of job displacement by race tend to be small and not exhibit strong patterns at different parts of the conditional distribution.  The effect of job displacement for college graduates is also very heterogeneous.  Some college graduates experience very large negative effects from job displacement, while college graduates (conditional on doing well following job displacement) tend to do better than high school graduates.  For individuals with less than a high school education, these results are almost exactly reversed.  Finally, our results indicate that being older increases the negative effects of job displacement on average.  Once again, though, the effects are quite heterogeneous -- among workers who are most negatively affected by job displacement, being older has an especially large negative effect.

These results suggest that the largest negative effects of job displacement tend to be for older, male, college graduates.  This should probably not be a surprising result as the effect of job displacement depends on both earnings following job displacement, $Y_{it}(1)$, as well as what earnings \textit{would have been} in the absence of job displacement, $Y_{it}(0)$.  Older, male, college graduates may have large earnings losses due to job displacement particularly in the case where if, had they not been displaced, they would have had high earnings.

\begin{figure}[t]
  \caption{Quantile Regression Estimates of Job Displacement Effects on $Y_t(0)$} \label{fig:5}
  \centering \includegraphics[width=.8\textwidth]{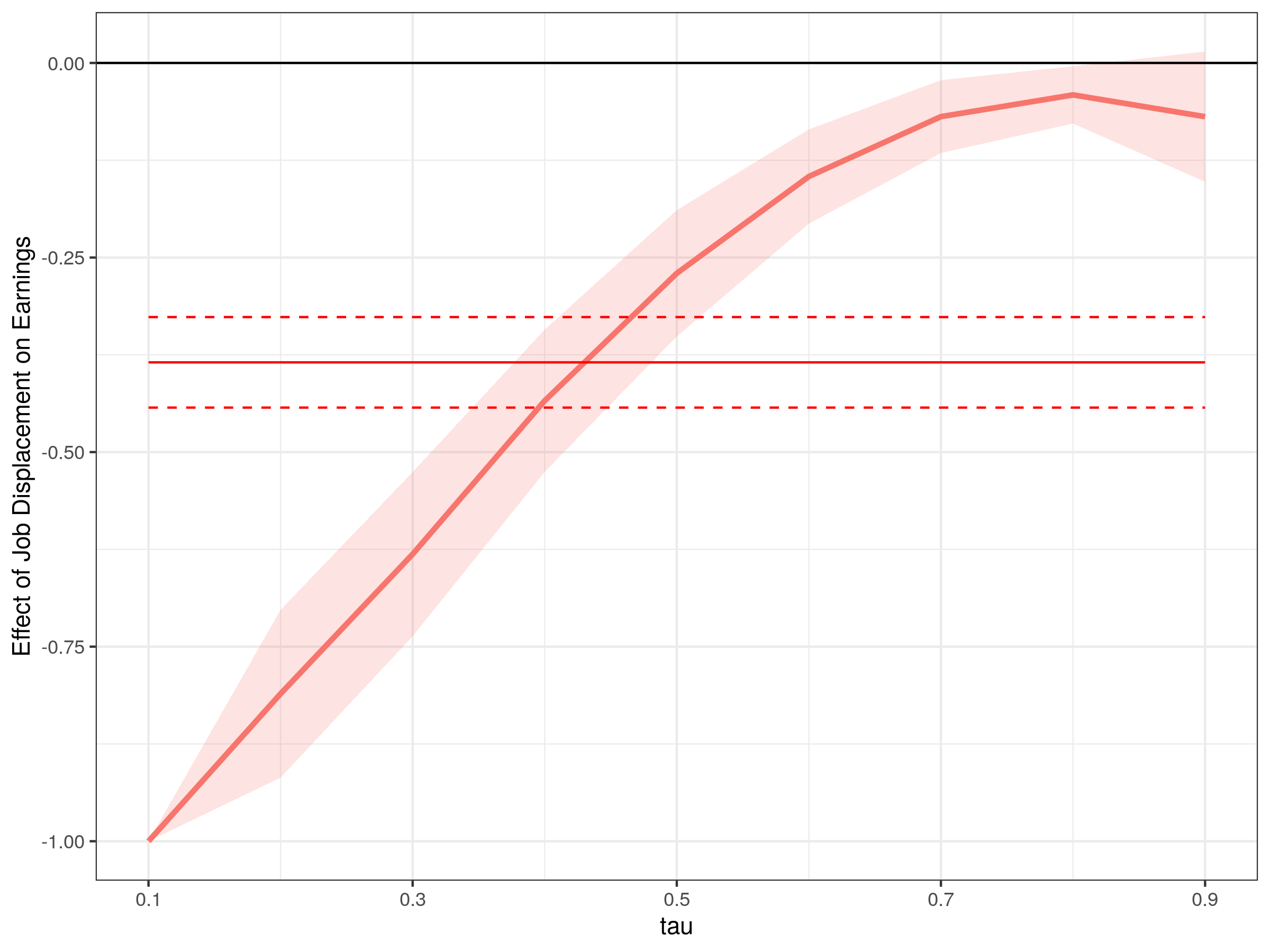}
  \subcaption*{\textit{Notes:} QR estimates of the effect of what earnings would have been in the absence of job displacement, $Y_t(0)$, on the quantiles of the effect of job displacement as discussed in the text.  The solid horizontal line provides OLS estimates of the effect of non-displaced potential earnings, and the horizontal dashed line contains a 90\% confidence interval.  The other line provides QR estimates of the effect of non-displaced potential earnings at particular quantiles from 0.1, 0.2, \ldots, 0.9.  The shaded area contains pointwise 90\% confidence intervals from the quantile regressions computed using the bootstrap with 1000 iterations.}
\end{figure}

Finally, in this section, we investigate how the effect of job displacement varies with $Y_t(0)$ -- i.e., how the effect varies across individuals as a function of what their earnings would have been if they had not been displaced.  We plot quantile regression estimates from regressing $(Y_{it}(1) - \hat{Y}_{it}(0))$ on $\hat{Y}_{it}(0)$ in \Cref{fig:5}.  On average, the effect of job displacement is larger for individuals who would have had higher earnings if they had not been displaced from their job and had maintained the same rank in the earnings distribution over time.  Notably, this is exactly the opposite interpretation as one would be tempted to make if one only compared the marginal distributions (see discussion in \Cref{rem:3}).  In addition, there is substantial heterogeneity.  For individuals that experience the largest negative effects of job displacement, these negative effects appear to be substantially larger for those who would have had high earnings if they had not been displaced and maintained their rank.  On the other hand, for those who are less affected by job displacement, the difference between the effect for those who would have had high earnings relative to those who would have had low earnings in the absence of job displacement is small.

\subsection{Results in Other Time Periods}

Next, we provide analogous results for 2009-2010 (corresponding to the Great Recession) and for 1997-1998 (an earlier time period when the U.S. economy was quite strong).  We briefly discuss the main results from these other periods below and provide additional details in the appendix (in particular, there we provide summary statistics and quantile regression estimates of the effect of covariates and the effect of non-displaced potential earnings on the distribution of the effect of job displacement).  All of the results in this section use the same identification arguments and estimation approaches that were applied to the 2015-2016 data above.

\begin{figure}[t] 
  \caption{Main Results for 2009-2010} \label{fig:2010-1}
  \includegraphics[width=0.5\textwidth]{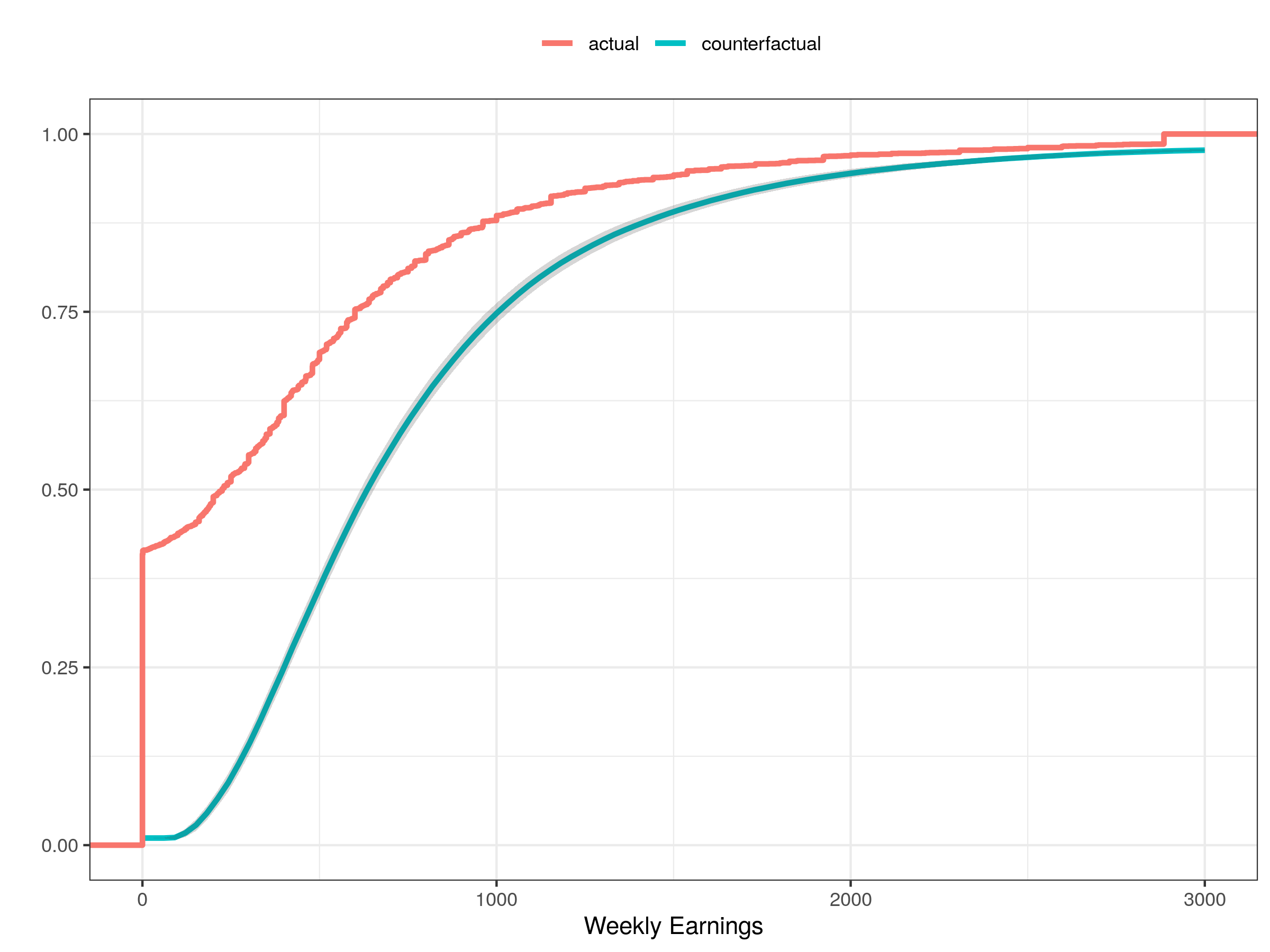}
  \includegraphics[width=0.46\textwidth]{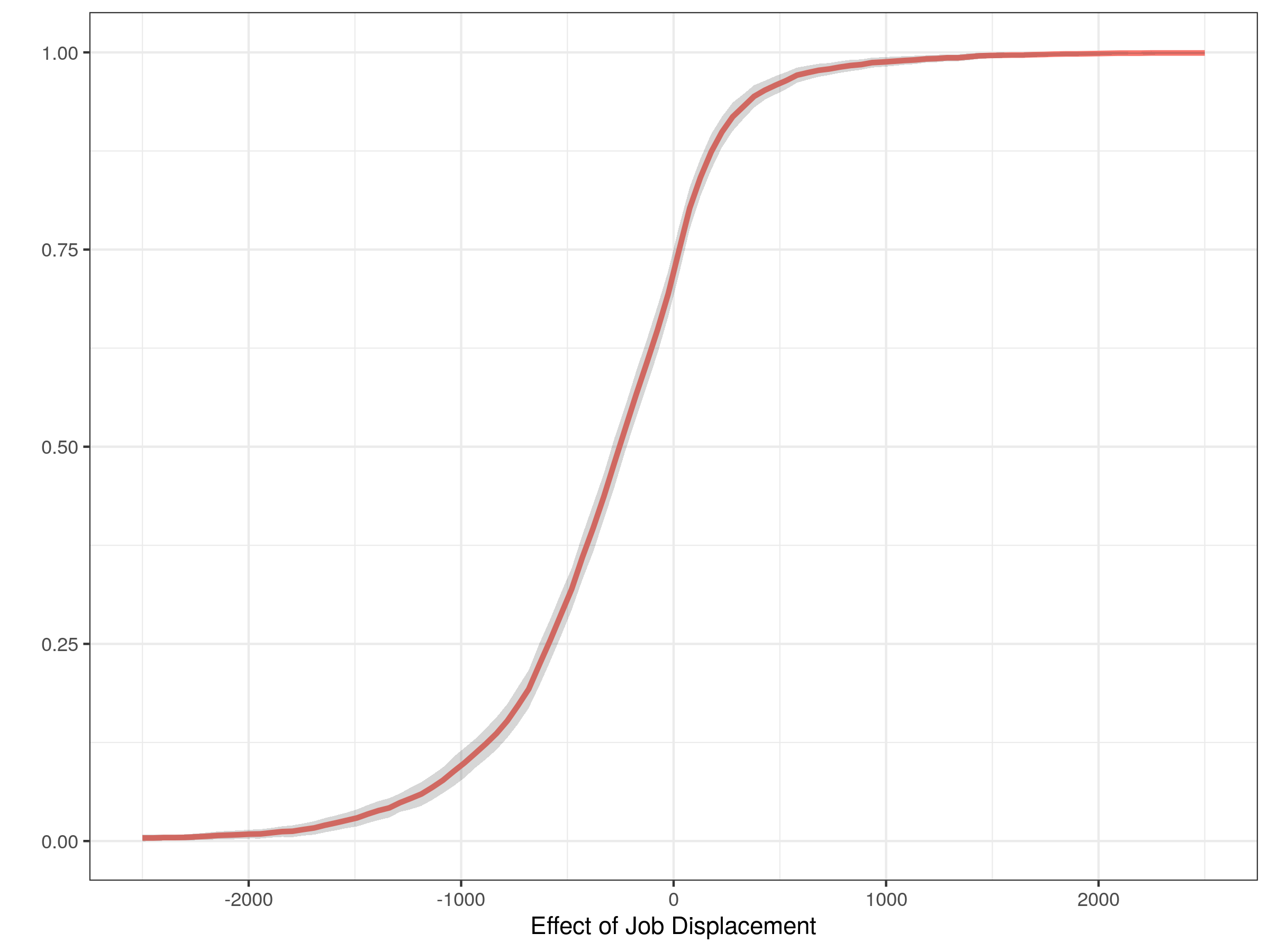}
  \subcaption*{\textit{Notes:}  The left panel contains plots of the distributions of earnings for displaced workers as well as their counterfactual distribution of earnings if they had not been displaced.  The right panel contains the distribution of the effect of job displacement for displaced workers as discussed in the text.}
\end{figure}

\paragraph{Results for 2009-2010} \

These results are for workers who were displaced at the height of the Great Recession.  \Cref{fig:2010-1} plots the observed and counterfactual distributions of outcomes for displaced workers.  The effects of job displacement are much larger, on average, than in 2015-2016.  We estimate that workers who were displaced from their job earn \$308 less on average than they would have earned if they had not been displaced; this corresponds to 39\% lower earnings on average relative to pre-displacement earnings.  Much of these effects appear to be driven by not being able to find work.  41\% of displaced workers continued to be unemployed in 2010 (for 2015-2016, only 20\% of displaced workers continued to be unemployed in 2016).  Interestingly, we estimate that 28\% of displaced workers have higher earnings than they would have had if they had not been displaced and maintained their rank in the earnings distribution over time.  Even when we just compare current earnings to pre-displacement earnings, we find that 21\% of displaced workers had higher earnings following displacement than they had in their pre-displacement job.  This indicates that even though the effects of job displacement appear to be substantially larger during the Great Recession, there is still substantial heterogeneity and a non-trivial fraction of displaced workers seem to have higher earnings than they otherwise would have had.

The effect of covariates on the distribution of the effect of job displacement in 2009-2010 is broadly similar to the effect in 2015-2016 (see \Cref{fig:2010-4} in the appendix).  In particular, among those who appear to be most negatively affected by job displacement, older, male, college graduates tend to be the most affected (here, there is also some evidence that among those most affected by job displacement, married displaced workers are also more negatively affected by job displacement).  Finally, similarly to 2015-2016, the effect of job displacement tends to be negatively related to earnings that individuals would have experienced if they had not been displaced (see \Cref{fig:2010-5} in the appendix).  These effects tend to be biggest for individuals who experience the largest negative effects of job displacement.  

\FloatBarrier

\paragraph{Results for 1997-1998} \

\begin{figure}[t] 
  \caption{Main Results for 1997-1998} \label{fig:1998-1}
  \includegraphics[width=0.5\textwidth]{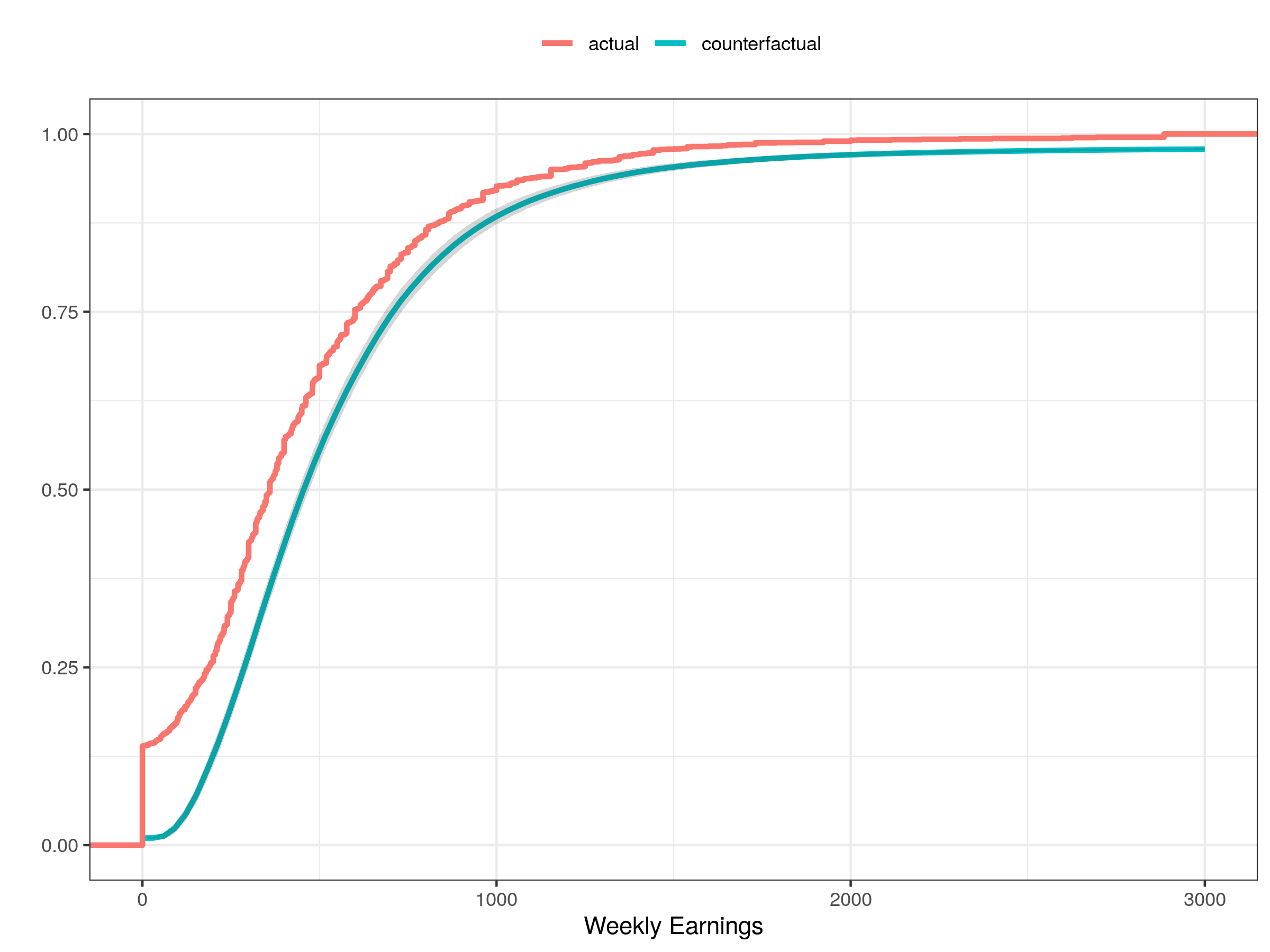}
  \includegraphics[width=0.46\textwidth]{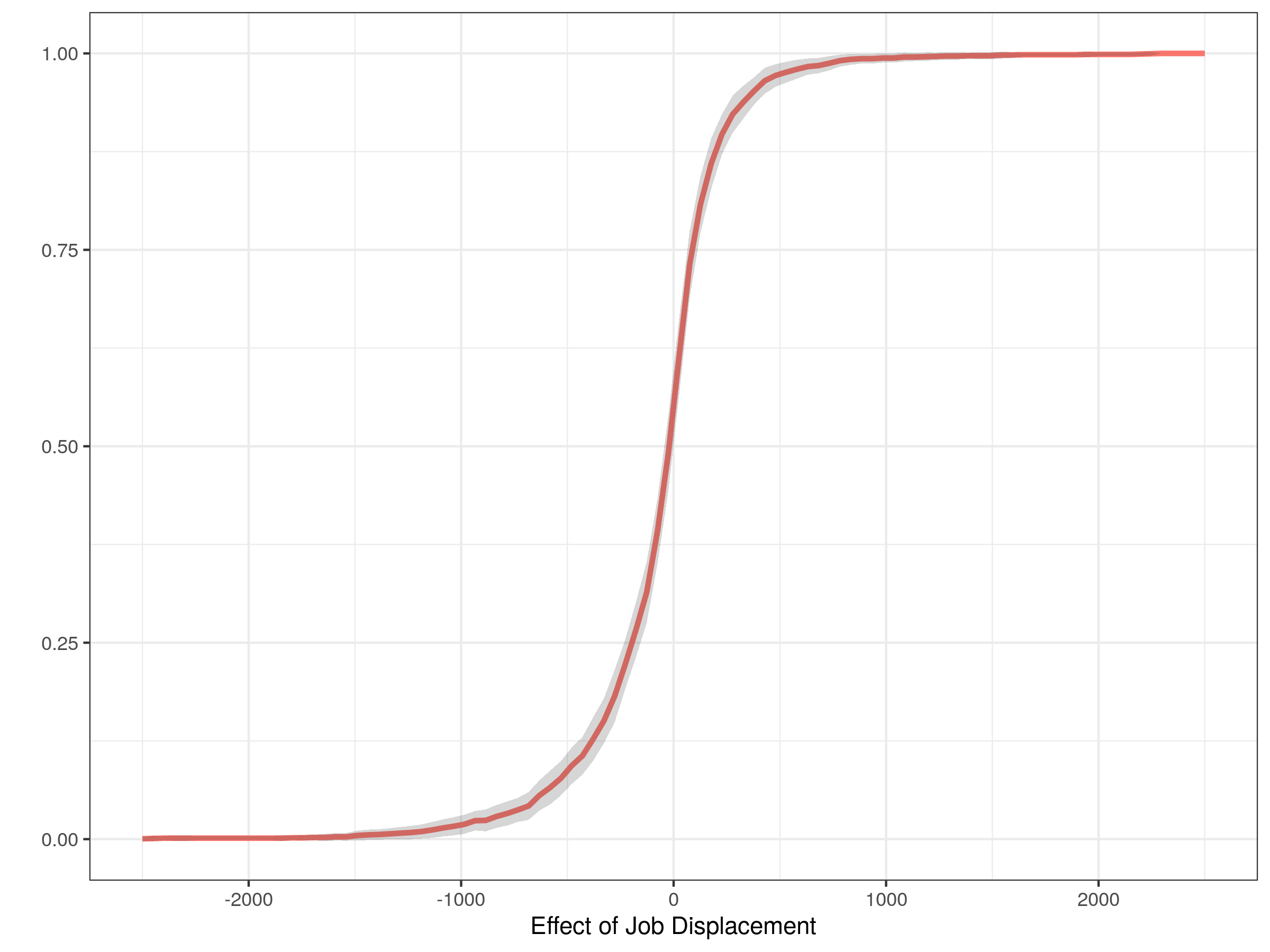}
  \subcaption*{\textit{Notes:}  The left panel contains plots of the distributions of earnings for displaced workers as well as their counterfactual distribution of earnings if they had not been displaced.  The right panel contains the distribution of the effect of job displacement for displaced workers as discussed in the text.}
\end{figure}

Finally, we examine the effect of job displacement in 1997-1998.  This is an earlier period than the ones we have considered so far, and the U.S. economy was quite strong during this period.  \Cref{fig:1998-1} contains our main estimates from this period.  We estimate that displaced workers earned on average \$68 less per week than they would have earned if they had not been displaced.  This estimate corresponds to 13\% lower earnings on average relative to pre-displacement earnings.  This estimate is somewhat smaller than the corresponding estimate for 2015-2016 and much smaller than the corresponding estimate for 2009-2010.  Part of the reason for the smaller effects here is that only 14\% of displaced workers are unemployed in 1998.  In addition, we estimate that 45\% of displaced workers have higher earnings than they would have had if they had not been displaced and maintained their rank in the earnings distribution over time.  Similarly, 42\% of displaced workers had higher earnings in their current job than they did in their pre-displacement job.  Finally, our estimates for the effect of covariates and untreated potential outcomes are broadly the same as in the other periods (see \Cref{fig:1998-4,fig:1998-5} in the appendix).  For displaced workers that are most affected by job displacement, these effects once again tend to be largest for older, male, college graduates and for those who would have had high earnings if they had not been displaced.  Given our results in earlier sections, these results seem to indicate that the finding that the largest effects of job displacement are concentrated among these groups is robust across time periods and across various states of the macroeconomy.

\FloatBarrier

\subsection{Robustness Checks}

To conclude this section, we discuss a number of additional robustness checks: (i) using Change in Changes as a replacement first step estimator, (ii) placebo tests related to our results on treatment effect heterogeneity, (iii) placebo tests related to regression towards mean earnings over time, and (iv) specification tests for our linear QR specifications.  In this section, for brevity, we discuss only the results for 2016.

\subsubsection*{Change in Changes}  
\begin{figure}
  \caption{Change in Changes as Alternative First-Step Estimator}
  \label{fig:cic}
  \begin{subfigure}{.5\textwidth}
    \centering \includegraphics[width=.9\textwidth]{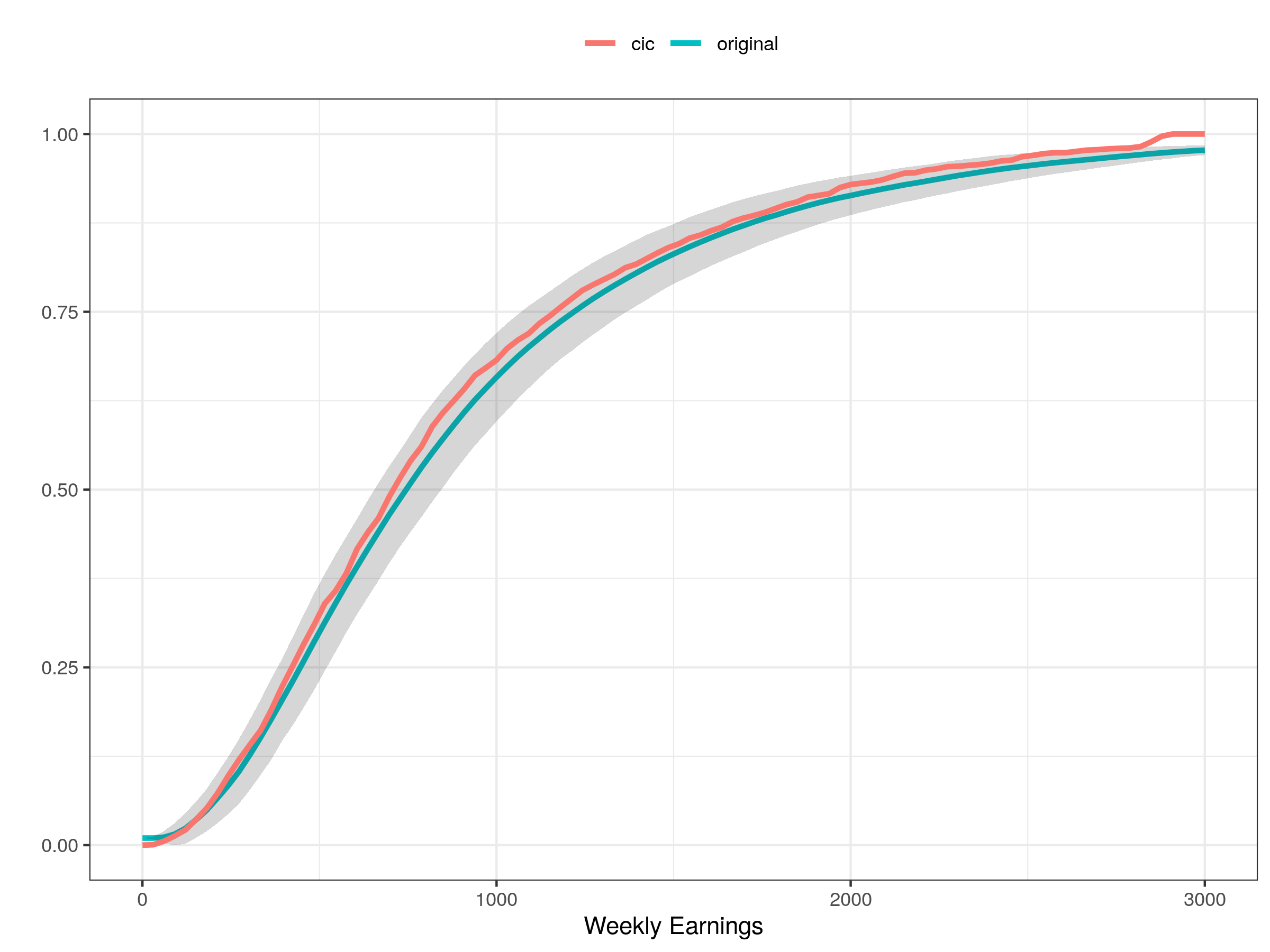}
    \caption{Counterfactual Distributions}
  \end{subfigure}
  \begin{subfigure}{.5\textwidth}
    \centering \includegraphics[width=.9\textwidth]{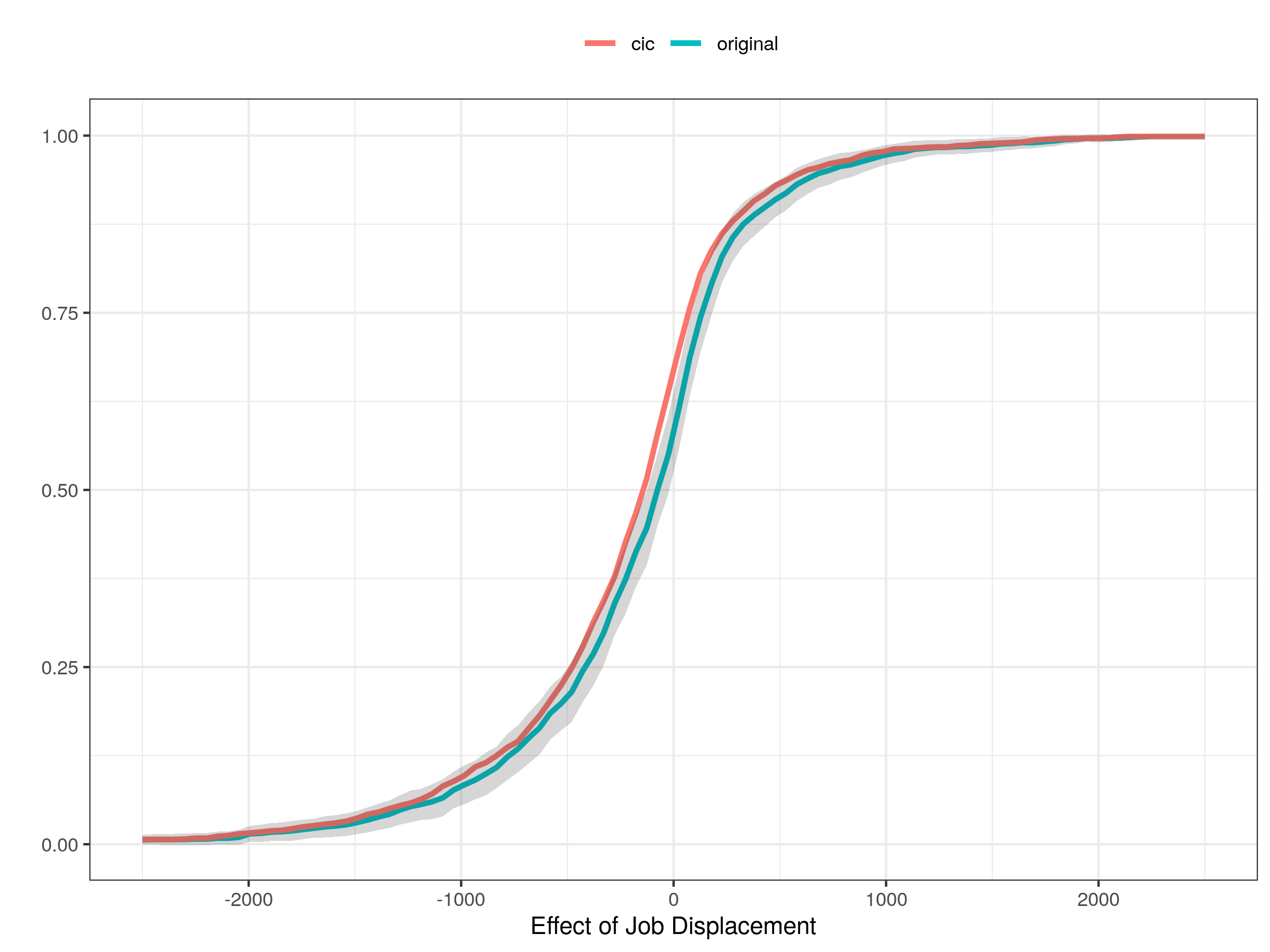}
    \caption{Distribution of the Effect of Job Displacement}
  \end{subfigure}
  \subcaption*{\textit{Notes:}  Panel (a) contains our original estimates of the counterfactual distribution of earnings that displaced workers would have experienced if they had not been displaced (which are available as the blue line along with a 95\% uniform confidence band) to the same counterfactual distribution that arises when Change in Changes is used as an alternative first-step estimator.  Panel (b) contains our original estimates of the distribution of the effect of job displacement (which are available as the blue line along with a 95\% uniform confidence band) along with estimates of the distribution of the effect of job displacement arising from using Change in Changes as an alternative first-step estimator.  The results in this figure are for 2016.}
\end{figure}

As discussed above, using Change in Changes (\citet{athey-imbens-2006,melly-santangelo-2015}) provides a reasonable alternative to our approach.  Change in Changes immediately recovers $F_{Y_t(0)|X,D=1}$ as in Step 1 of our identification argument.  Then, the same arguments as in our Step 2 immediately apply.  Panel (a) of \Cref{fig:cic} provides our original estimate of the counterfactual distribution along with the corresponding estimate when Change in Changes is used in the first-step instead.  These results are very similar.

Likewise, Panel (b) of \Cref{fig:cic} plots estimates of the distribution of the effect of job displacement in both of these cases.\footnote{For the distribution of the effect of job displacement, there are some places where the estimate of this distribution using Change in Changes does fall outside of the 95\% uniform confidence band using our original approach.  However, despite the uniform confidence bands, the figure does not provide a formal test that the distributions are equal because it does not account for sampling uncertainty using Change in Changes.}  These results are also quite similar; the results coming from Change in Changes generally indicate a somewhat more negative effect of job displacement.  On average, we originally estimate that displaced workers lose \$157 per week relative to what they would have earned if they had not been displaced; using Change in Changes, we estimate that they would have lost \$227 per week.  We originally estimate that workers in the 10th percentile of the effect of job displacement (these are among those most negatively affected by job displacement) lose \$881 per week; using Change in Changes, we estimate that they lose \$957 per week.  For workers in the 90th percentile of the effect of job displacement, we originally estimate that their earnings increase by \$436 per week; using Change in Changes, we estimate that they earn \$347 more.

\subsubsection*{Placebo Tests for Treatment Effect Heterogeneity}

Our main results on treatment effect heterogeneity rely heavily on rank invariance as in \Cref{ass:riot}.  Recall that, under our other identifying assumptions, for displaced workers, we identify $\tilde{Y}_{it}(0)$ -- this is the outcome individual $i$ would experience if they had not been displaced and had the same rank in the earnings distribution as in the previous period.  Rank invariance additionally implies that $Y_{it}(0) = \tilde{Y}_{it}(0)$.  It is possible that treatment effect heterogeneity could arise spuriously due to violations of rank invariance.  To see this, consider the case where there is no effect of job displacement on earnings so that $Y_{it}(1) = Y_{it}(0)$ for all individuals.  Even in this case, if rank invariance does not hold, then our approach would still detect some heterogeneous effects of job displacement (since $Y_{it}(1) \neq \tilde{Y}_{it}(0)$).

Next, we discuss several ways to detect if there appears to be meaningful heterogeneity apart from rank invariance.  First, we compute Spearman's Rho (the rank correlation) for $Y_t$ and $Y_{t-1}$ for both the group of displaced and non-displaced workers.  Rank invariance holds if Spearman's Rho is equal to 1.  For non-displaced workers, Spearman's Rho is equal to 0.69.  This is strong positive dependence but, since it is less than 1, it essentially implies that rank invariance does not hold for non-displaced workers.\footnote{This does imply that \textit{unconditional} rank invariance does not hold here, but our assumptions have been about conditional rank invariance which is somewhat weaker.}  One way to check if there is meaningful heterogeneity is to compare Spearman's Rho for displaced and non-displaced workers.  If the correlation of ranks of earnings is similar for displaced workers and non-displaced workers, this would be a piece of evidence against meaningful heterogeneity due to job displacement.  Instead, for displaced workers, Spearman's Rho is 0.44.  This is substantially lower; it means that displaced workers ranks in the earnings distribution are less correlated over time than non-displaced workers.  And it is suggestive evidence that there is additional heterogeneity beyond that mechanically introduced by violations of rank invariance.

Second, we propose to impute $Y_{it}(0)$ for \textit{non-displaced} individuals under the assumption of rank invariance; in particular, we impute $\ddot{Y}_{it}(0) := Q_{Y_t|X,D=0}(F_{Y_{t-1}|X,D=0}(Y_{it-1}))$ which is the earnings of non-displaced individual $i$ in the second time period if they maintained their rank from the first period.  If rank invariance holds for the group of non-displaced workers, then $Y_{it}(0)$ (which is observed for non-displaced workers) should be equal to $\ddot{Y}_{it}(0)$.  We compute the standard deviation of $Y_{it}(1) - \hat{Y}_{it}(0)$ for the group of displaced workers and $Y_t(0) - \hat{\ddot{Y}}_{it}(0)$ for the group of non-displaced workers (where $\hat{\ddot{Y}}_{it}(0)$ is a non-displaced workers earnings imputed as above with conditional distributions/quantiles estimated by quantile regression).  For non-displaced workers, the standard deviation is \$515, and for displaced workers, the standard deviation is \$617 -- once again suggesting that there is some additional heterogeneity due to job displacement that is not directly coming from the assumption of rank invariance.

\begin{figure}[t]
  \caption{Placebo Treatment Effect Heterogeneity}
  \label{fig:placebo-het}
  \centering \includegraphics[width=0.8\textwidth]{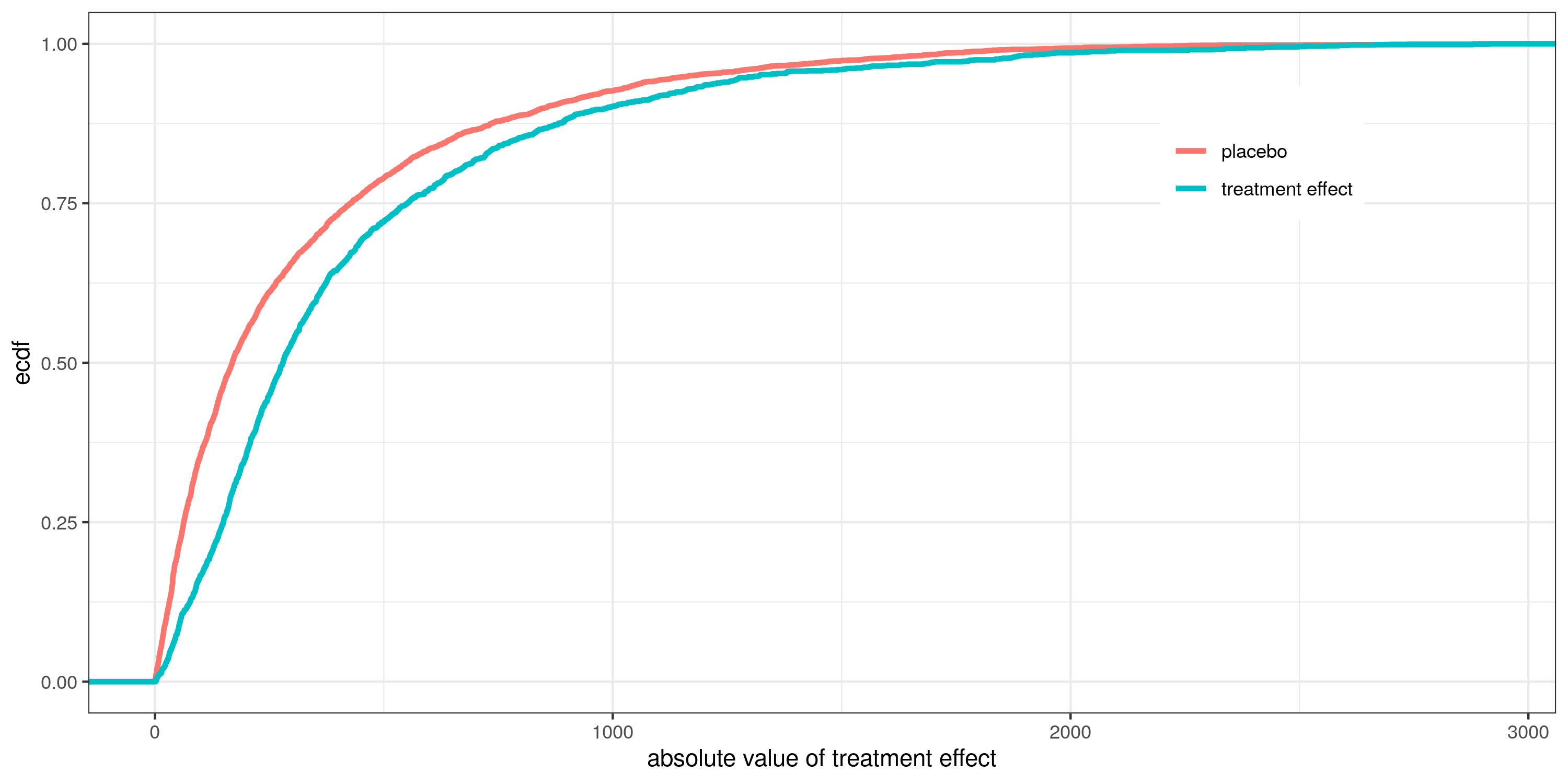}
  \subcaption*{\textit{Notes:} Plots of treatment effect heterogeneity under rank invariance for the group of displaced workers (blue line) and placebo treatment effect heterogeneity under rank invariance for the group of non-displaced workers (red line) as discussed in the text.}
\end{figure}

Finally, we plot the distribution of $| Y_{it}(1) - \hat{Y}_{it}(0) - \hat{\mu}_1|$ (where $\hat{\mu}_1$ is the average difference between $Y_{it}(1)$ and $\hat{Y}_{it}(0)$) for the group of displaced workers as well as the analogous distribution of placebo effects for the group of non-displaced workers.  These results are provided in \Cref{fig:placebo-het} and are similar to the previous ones -- there appears to be more treatment effect heterogeneity for displaced workers than what arises for non-displaced workers when imputing their earnings under rank invariance.  Taken together, these results suggest that, while our results can partially be explained by ``spurious'' heterogeneity arising from violations of the assumption of rank invariance, there still appears to be meaningful heterogeneity apart from rank invariance.

\subsubsection*{Placebo Tests for Regression to the Mean}

Another of our main results was that the effect of job displacement was larger for individuals who would have had higher earnings in the absence of job displacement relative to individuals that would have had lower earnings (see \Cref{fig:5} and related discussion).  This result also relied on the rank invariance condition in \Cref{ass:riot}.  An alternative explanation for this sort of result is that there is regression to the mean -- i.e., that individuals with higher earnings would have tended to experience relative decreases in their earnings even in the absence of job displacement.  Regression to the mean implies a particular violation of rank invariance that could potentially lead to results that look like those in \Cref{fig:5} even if the effect of job displacement does not systematically vary with earnings in the absence of displacement.

For example, suppose that the effect of job displacement is constant, i.e., $Y_{it}(1) - Y_{it}(0) = \alpha$ for all $i$.  Violations of rank invariance mean that $\tilde{Y}_{it}(0) \neq Y_{it}(0)$, but, in addition, regression to the mean says that $Y_{it}(0)$ will tend to be less than $\tilde{Y}_{it}(0)$ for large values of $\tilde{Y}_{it}(0)$ and will tend to be greater than $\tilde{Y}_{it}(0)$ for small values of $\tilde{Y}_{it}(0)$.  In other words, imposing rank invariance in the presence of regression to the mean tends to overstate earnings of high earners and understate earnings of low earners (both in the absence of displacement).  This suggests that regression to the mean could lead to results that are quantitatively similar to the ones that we report.

\begin{figure}[t]
  \caption{Placebo Tests for Regression to the Mean}
  \label{fig:placebo-rtm}
  \centering \includegraphics[width=.8\textwidth]{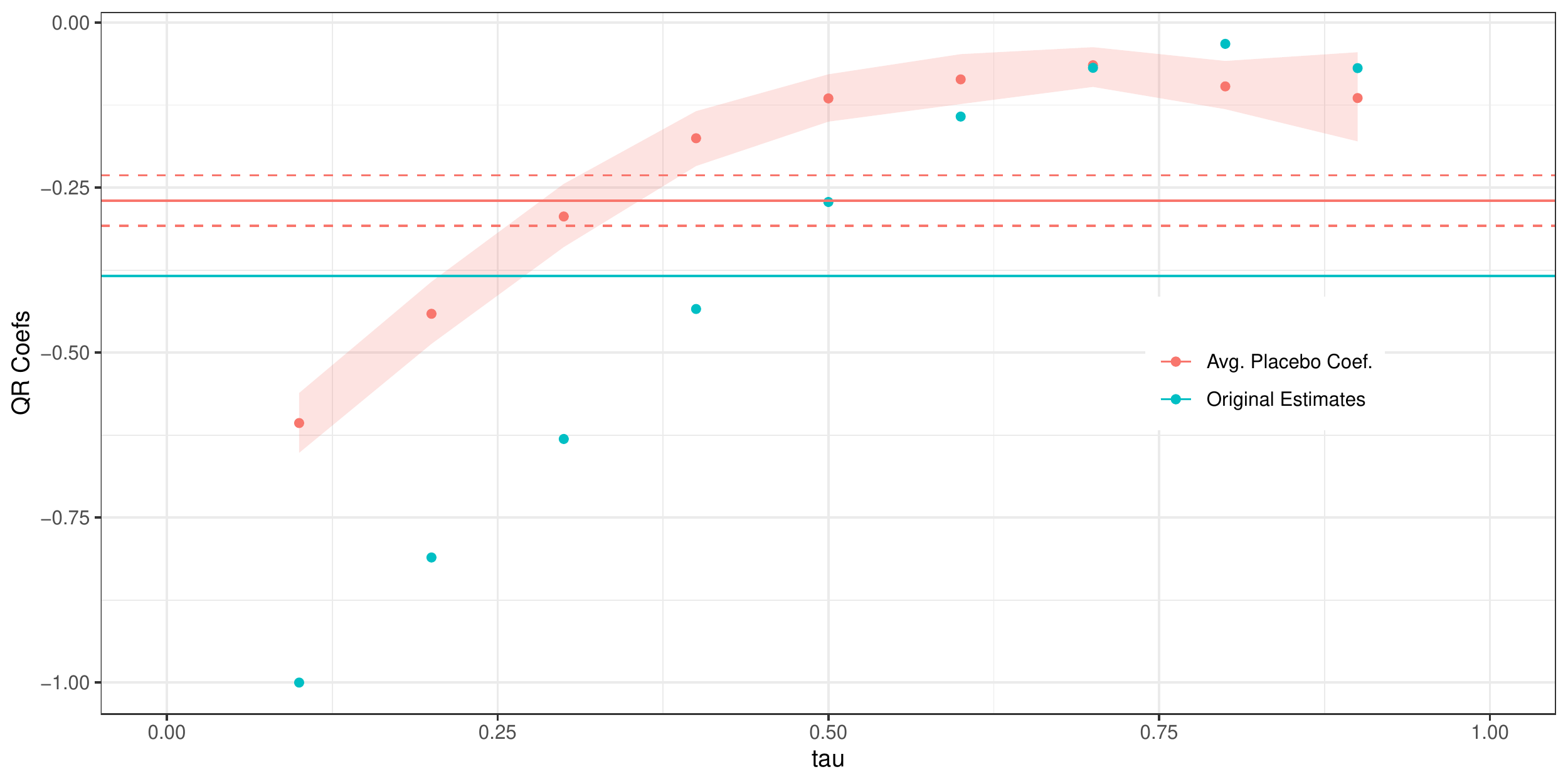}
  
  \subcaption*{\textit{Notes:} The red dots are the average quantile regression coefficients on earnings in the absence of job displacement  over 1000 placebo estimates using the approach discussed in the text.  The confidence bands are the corresponding 95th and 5th percentiles of the quantile regression coefficients.  The red line is the average OLS estimate of the effect of earnings in the absence of job displacement over 1000 placebo estimates, and the dotted red lines are the corresponding 95th and 5th percentiles.  The blue dots and blue line provide our original corresponding estimates of the quantile regression and OLS estimates.  These results are for 2016.}
\end{figure}

In order to think about regression to the mean, we once again exploit having access to the untreated group.  Our goal here is to think about whether or not our results on differential effects of job displacement by earnings in the absence of displacement can be explained by regression to the mean of earnings.  In order to address this, we take the group of non-displaced workers and randomly assign 1633 (this is the original number of displaced workers) to be treated.  We assign the remaining individuals to be untreated.  By construction, the ``effect'' of job displacement should not vary across different values of earnings in the absence of displacement.\footnote{Note that, besides \Cref{ass:riot}, our other identifying assumptions hold by construction in this setup.}

We construct this sort of dataset 1000 times and, for each one, we compute the same estimates (QR and OLS) of job displacement effects on earnings in the absence of displacement (i.e., we replicate the analysis in Figure 5 but with an adjusted dataset).  These results are reported in \Cref{fig:placebo-rtm}.  When we use this approach, we find that placebo-displaced workers tend to experience larger placebo-declines in earnings across higher imputed earnings -- this is consistent with the idea that, even in the absence of job displacement, there would be some regression to the mean in terms of earnings.  That being said, our original estimates are noticeably larger in magnitude.  Together, these results suggest that our original results may be somewhat biased (due to regression to the mean), but that we are still picking up a meaningful pattern that the effects of displacement appear to be larger for higher earning individuals.
    
\subsubsection*{Quantile Regression Specification Tests}

{ \setlength{\tabcolsep}{30pt}
\ctable[caption = {\citet{rothe-wied-2013} QR specification test},pos=t,label=tab:rw]{rlc}{\tnote[]{\textit{Notes:} Specification tests for the linear quantile regression specifications for $Y_{t-1}$ in \Cref{ass:qr}.  The reported p-value is calculated using 100 bootstrap iterations.}}{\FL
  year & group         & p-value \ML
  2016 & displaced     & 0.24 \NN
  2016 & non-displaced & 0.58 \NN
  2010 & displaced     & 0.71 \NN
  2010 & non-displaced & 1.00 \NN
  1998 & displaced     & 0.73 \NN
  1998 & non-displaced & 0.67 \LL
}
}

We have used quantile regression throughout the paper to estimate conditional distributions and conditional quantiles.  As discussed above, using quantile regression in our application has a number of advantages and nicely balances flexibility of the estimator with feasibility in our setup.  That being said, the quantile regression specifications specifications do involve parametric assumptions, and these assumptions are testable.  We implemented the specification test proposed in \citet{rothe-wied-2013}.  Here, we focus on testing the specifications in \Cref{ass:qr} only for earnings in period $t-1$ conditional on covariates and separately for displaced and non-displaced workers.\footnote{These are the quantile regressions that show up in the first step of our estimation procedure.  %
  We do use quantile regression in second step estimates, but it is not immediately clear how to adapt the \citet{rothe-wied-2013} test to that case due to there not being an observed distribution that we can directly compare to.}  We report these results in \Cref{tab:rw}.  We do not reject the linear quantile regression specification for any of our estimators.

\section{Conclusion }

In this paper, we have studied how the effect of job displacement varies across different individuals.  To do this, we compared the earnings of displaced workers to what their earnings would have been if they had maintained their rank in the (conditional) distribution of weekly earnings (though we allow for the distribution of weekly earnings to change over time).  A key requirement of our approach was to be able to estimate several conditional distributions in a first step in a way that was both not too restrictive while also being feasible with a moderate sized data set and a fairly large number of covariates.  To do this, we relied heavily on first-step quantile regression estimators.

We found that displaced workers lose about \$157 per week due to job displacement, on average.  However, that average effect masks a tremendous amount of heterogeneity.  A large fraction of workers appear to be no worse off (or even better off) in terms of their earnings following job displacement than they would have been if they had not been displaced.  On the other hand, another large fraction loses substantially more than the average earnings loss.  Once we had obtained the distribution of the effect of job displacement, we used quantile regression to study how the distribution of the effect of job displacement depends on covariates.  We also showed that this distribution varies substantially across sex, education levels, and age, as well as across different amounts of earnings that displaced workers would have had if they had not been displaced.

\newpage

\clearpage

\appendix

\onehalfspacing

\section*{Compliance with Ethical Standards}

\noindent Conflict of Interest:  Afrouz Azadikhah Jahromi declares that she has no conflict of interest.  Brantly Callaway declares that he has no conflict of interest.

\bigskip

\noindent Ethical approval: This article does not contain any studies with human participants or animals performed by any of the authors.

\printbibliography

\pagebreak

\section{Additional Tables and Figures}

\FloatBarrier

\subsection{2009-2010 Results}
\newcolumntype{.}{D{.}{.}{-1}}
\ctable[caption={2009-2010 Summary Statistics},label=,pos=!h,]{lrrrr}{\tnote[]{\textit{Sources: CPS and Displaced Workers Survey}}}{\FL
\multicolumn{1}{l}{}&\multicolumn{1}{c}{Displaced}&\multicolumn{1}{c}{Non-Displaced}&\multicolumn{1}{c}{Difference}&\multicolumn{1}{c}{P-val on Difference}\ML
{\bfseries Earnings}&&&&\NN
~~2010 Earnings&418.37&887.54&-469.169&0.00\NN
~~2009 Earnings&797.24&876.66&-79.424&0.00\NN
~~Change Earnings&-378.87&10.87&-389.745&0.00\ML
{\bfseries Covariates}&&&&\NN
~~Male&0.60&0.48&0.122&0.00\NN
~~White&0.83&0.85&-0.017&0.04\NN
~~Married&0.53&0.62&-0.096&0.00\NN
~~College&0.25&0.38&-0.123&0.00\NN
~~Less HS&0.10&0.06&0.044&0.00\NN
~~Age&41.50&44.87&-3.371&0.00\NN
~~N&4088&4000&&\LL
}

\begin{figure}[t!]
  \caption{2009-2010 Quantile Regression Estimates of Job Displacement Effects on Covariates} \label{fig:2010-4}
  \centering \includegraphics[width=\textwidth]{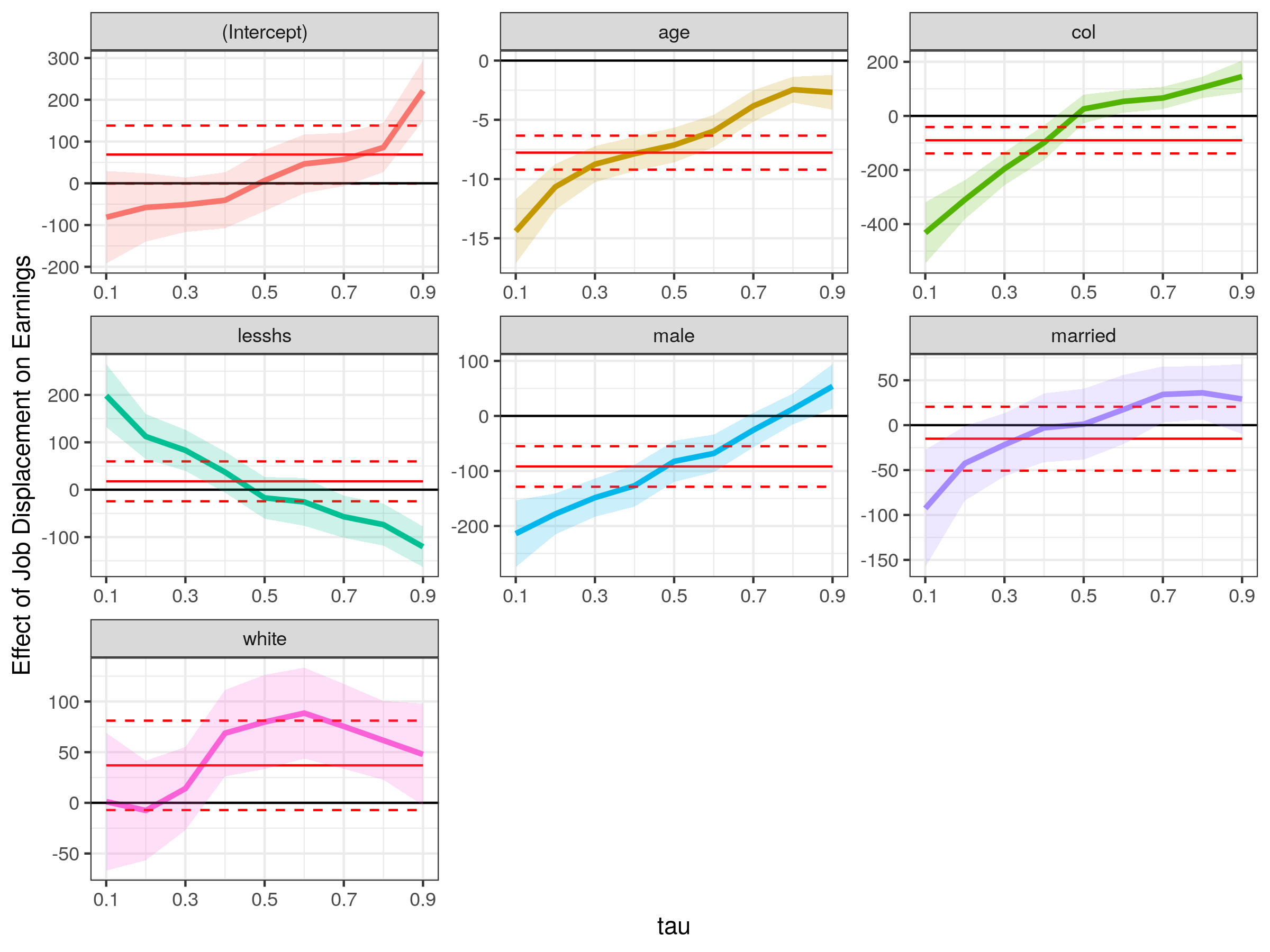}
  \subcaption*{\textit{Notes:} QR estimates of the effect of covariates on the quantiles of the effect of job displacement in 2009-2010.  The solid red horizontal lines provide OLS estimates of the effect of each covariate, and the horizontal dashed line contains a 90\% confidence interval.  The other multi-colored lines provide QR estimates of the effect of each covariate at particular quantiles from 0.1, 0.2, \ldots, 0.9.  The shaded areas contain pointwise 90\% confidence intervals from the quantile regressions.}
\end{figure}

\begin{figure}[t]
  \caption{2009-2010 Quantile Regression Estimates of Job Displacement Effects on $Y_t(0)$} \label{fig:2010-5}
  \centering \includegraphics[width=.8\textwidth]{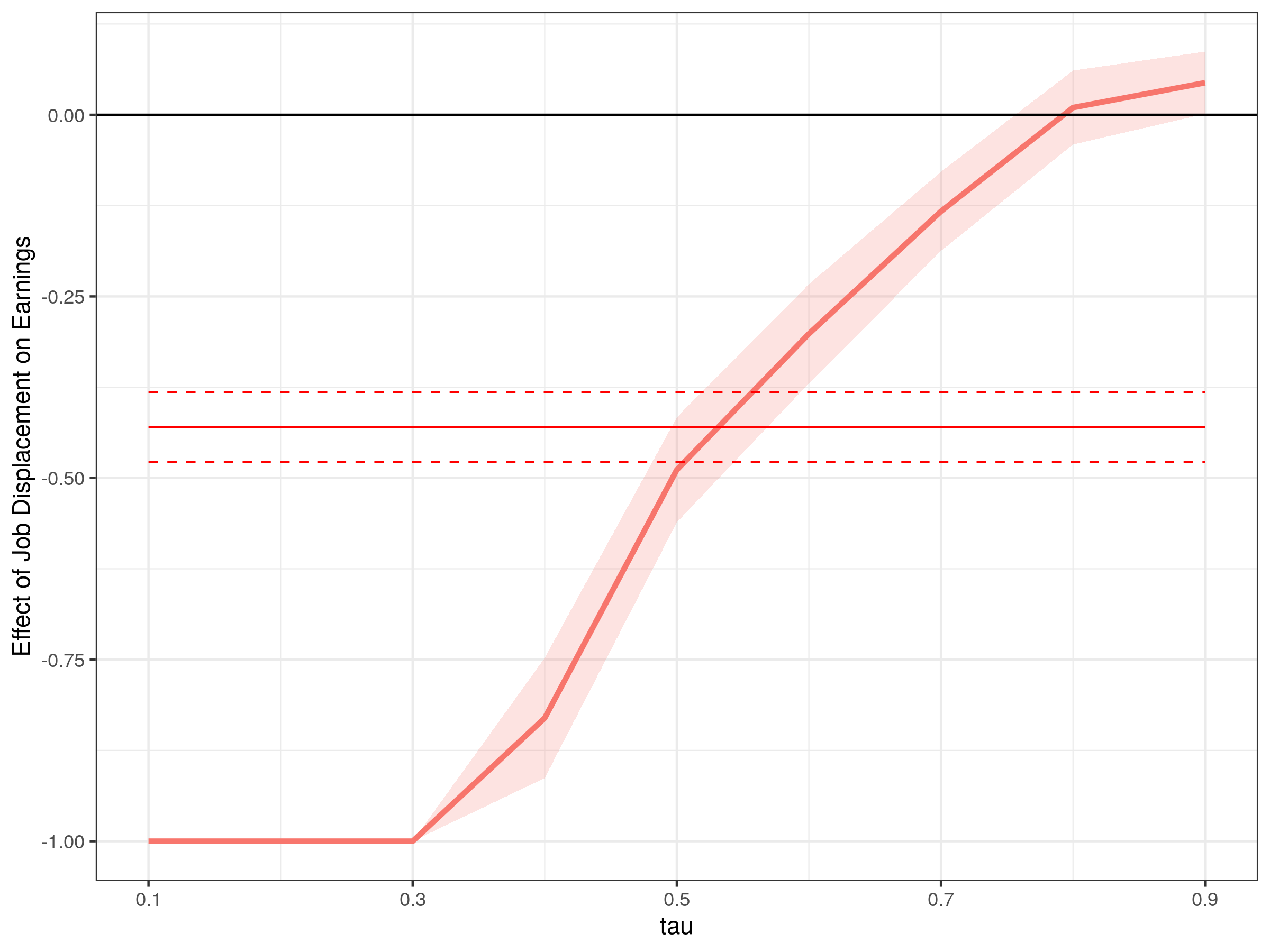}
  \subcaption*{\textit{Notes:} QR estimates of the effect of what earnings would have been in the absence of job displacement, $Y_t(0)$, on the quantiles of the effect of job displacement as discussed in the text in 2009-2010.  The solid horizontal line provides OLS estimates of the effect of non-displaced potential earnings, and the horizontal dashed line contains a 90\% confidence interval.  The other line provides QR estimates of the effect of non-displaced potential earnings at particular quantiles from 0.1, 0.2, \ldots, 0.9.  The shaded area contains pointwise 90\% confidence intervals from the quantile regressions computed using the bootstrap with 1000 iterations.}
\end{figure}

\FloatBarrier

\subsection{1997-1998 Results}

\newcolumntype{.}{D{.}{.}{-1}}
\ctable[caption={1997-1998 Summary Statistics},label=,pos=!h,]{lrrrr}{\tnote[]{\textit{Sources: CPS and Displaced Workers Survey}}}{\FL
\multicolumn{1}{l}{}&\multicolumn{1}{c}{Displaced}&\multicolumn{1}{c}{Non-Displaced}&\multicolumn{1}{c}{Difference}&\multicolumn{1}{c}{P-val on Difference}\ML
{\bfseries Earnings}&&&&\NN
~~1998 Earnings&443.38&621.07&-177.692&0.00\NN
~~1997 Earnings&526.52&582.40&-55.886&0.00\NN
~~Change Earnings&-83.14&38.67&-121.806&0.00\ML
{\bfseries Covariates}&&&&\NN
~~Male&0.56&0.50&0.057&0.00\NN
~~White&0.87&0.88&-0.008&0.34\NN
~~Married&0.55&0.66&-0.108&0.00\NN
~~College&0.24&0.31&-0.069&0.00\NN
~~Less HS&0.12&0.08&0.038&0.00\NN
~~Age&38.69&42.03&-3.34&0.00\NN
~~N&2387&3842&&\LL
}

\begin{figure}[t!]
  \caption{1997-1998 Quantile Regression Estimates of Job Displacement Effects on Covariates} \label{fig:1998-4}
  \centering \includegraphics[width=.9\textwidth]{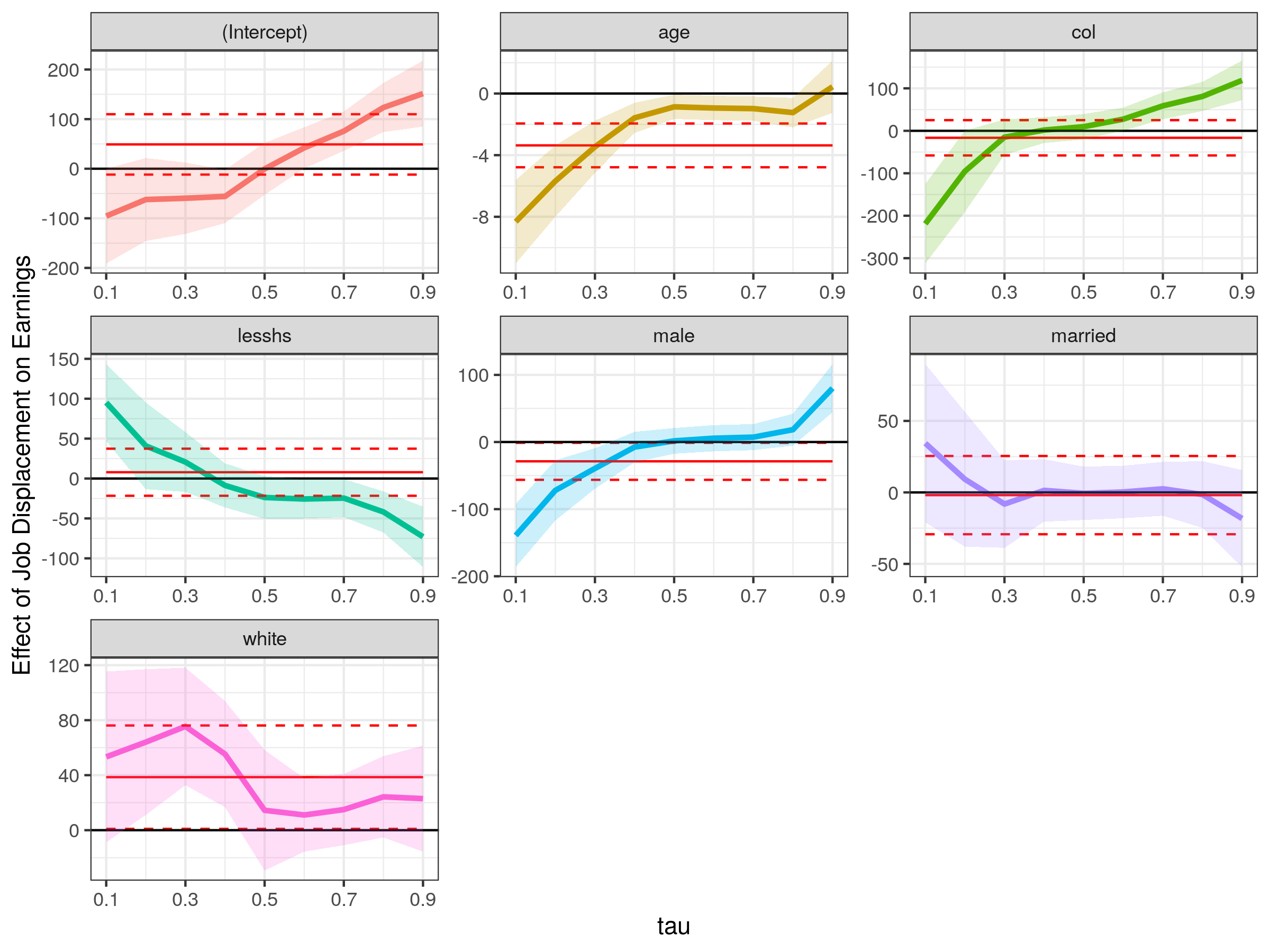}
  \subcaption*{\textit{Notes:} QR estimates of the effect of covariates on the quantiles of the effect of job displacement in 1997-1998.  The solid red horizontal lines provide OLS estimates of the effect of each covariate, and the horizontal dashed line contains a 90\% confidence interval.  The other multi-colored lines provide QR estimates of the effect of each covariate at particular quantiles from 0.1, 0.2, \ldots, 0.9.  The shaded areas contain pointwise 90\% confidence intervals from the quantile regressions.}
\end{figure}

\begin{figure}[t]
  \caption{1997-1998 Quantile Regression Estimates of Job Displacement Effects $Y_t(0)$} \label{fig:1998-5}
  \centering \includegraphics[width=.8\textwidth]{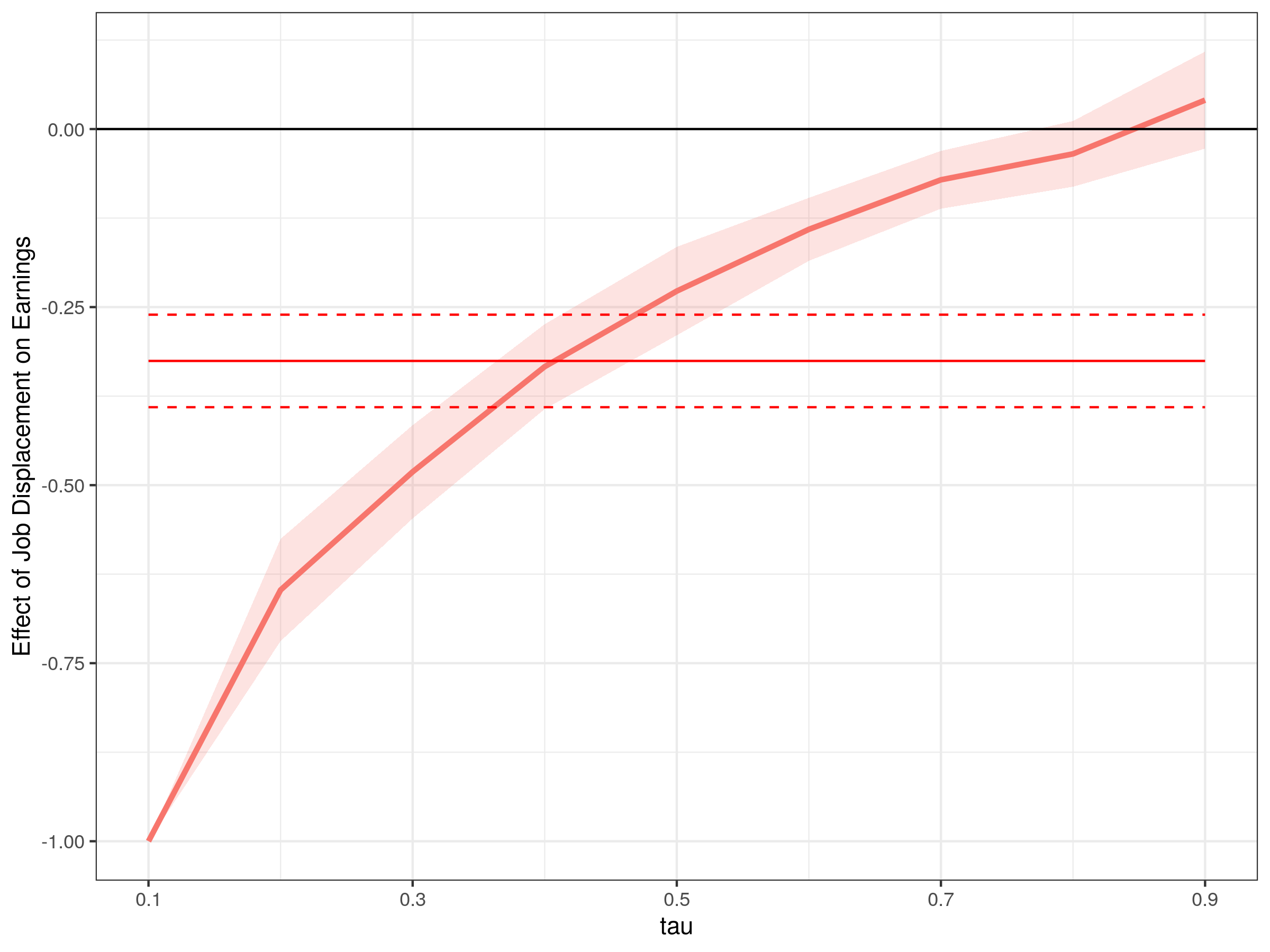}
  \subcaption*{\textit{Notes:} QR estimates of the effect of what earnings would have been in the absence of job displacement, $Y_t(0)$, on the quantiles of the effect of job displacement as discussed in the text in 1997-1998.  The solid horizontal line provides OLS estimates of the effect of non-displaced potential earnings, and the horizontal dashed line contains a 90\% confidence interval.  The other line provides QR estimates of the effect of non-displaced potential earnings at particular quantiles from 0.1, 0.2, \ldots, 0.9.  The shaded area contains pointwise 90\% confidence intervals from the quantile regressions computed using the bootstrap with 1000 iterations.}
\end{figure}

\end{document}